\documentclass[11pt]{article}

\newsavebox{\foobox}
\newcommand{\slantbox}[2][0]{\mbox{%
        \sbox{\foobox}{#2}%
        \hskip\wd\foobox
        \pdfsave
        \pdfsetmatrix{1 0 #1 1}%
        \llap{\usebox{\foobox}}%
        \pdfrestore
}}
\newcommand\unslant[2][-.25]{\slantbox[#1]{$#2$}}

\newcommand{\mpi}{\text{\unslant[-.18]\pi}}
\newcommand{\mdelta}{\text{\unslant[-.18]\delta}}

\renewcommand\textfraction{.05}

\usepackage[left=2cm, right=2cm, top=2.5cm, bottom=2.5cm]{geometry}
\usepackage[x11names]{xcolor}
\usepackage{fancyhdr, amssymb, cancel, amsmath, graphicx, pgfplots, tikz}
\usepackage{isomath}

\usetikzlibrary{shadows}

\newcommand{\stylecolor}{blue!50!black}

\usepackage[labelfont={bf,sf, color=\stylecolor}, margin={1.5cm,0cm}]{caption}

\usepackage[colorlinks=true, urlcolor=\stylecolor!70!white, linkcolor=\stylecolor, citecolor=\stylecolor!70!white, hyperindex=true, linktocpage=true]{hyperref}

\usepackage[explicit]{titlesec}

\newcommand*\sectionlabel{}
\titleformat{\section}
  {\gdef\sectionlabel{}
   \Large\bfseries\scshape}
  {\gdef\sectionlabel{\thesection }}{0pt}
  {\begin{tikzpicture}[remember picture,overlay]
       \end{tikzpicture}
  }
\titlespacing*{\section}{0pt}{0pt}{0pt}

\newcommand*\subsectionlabel{}
\titleformat{\subsection}
  {\gdef\subsectionlabel{}
   \large\bfseries\scshape}
  {\gdef\subsectionlabel{\thesubsection  }}{0pt}
  {\begin{tikzpicture}[remember picture,overlay]
    	\draw (-0.15, 0.02) node[right] {\color{\stylecolor} \textsf{\subsectionlabel}};
	\draw (1.25, 0) node[right] {\color{\stylecolor} \textsf{#1}};
	\fill[color=\stylecolor] (1,-0.25) rectangle (1.1, 0.25);
       \end{tikzpicture}
  }
\titlespacing*{\subsection}{0pt}{10pt}{10pt}

\newcommand*\subsubsectionlabel{}
\titleformat{\subsubsection}
  {\gdef\subsubsectionlabel{}
   \bfseries\scshape}
  {\gdef\subsubsectionlabel{\thesubsubsection.\ \  }}{0pt}
  {\begin{tikzpicture}[remember picture,overlay]
    	\draw (-0.15, 0) node[right] {\color{\stylecolor} \textsf{\subsubsectionlabel#1}};
       \end{tikzpicture}
  }
\titlespacing*{\subsubsection}{0pt}{7pt}{7pt}

\pgfplotsset{every axis legend/.append style={at={(1.02,1)},anchor=north west}}

\begin{document}

\allowdisplaybreaks

\renewcommand{\labelitemi}{\color{\stylecolor} $\blacktriangleright$}

\setcounter{totalnumber}{5}
\renewcommand\textfraction{.1}

\pagestyle{fancy}
\renewcommand{\headrulewidth}{0pt}
\fancyhead{}

\fancyfoot{}
\fancyfoot[C] {\textsf{\textbf{\thepage}}}

\begin{equation*}
\begin{tikzpicture}
\draw (\textwidth, 0) node[text width = \textwidth, right] {\color{white} easter egg};
\end{tikzpicture}
\end{equation*}

\begin{equation*}
\begin{tikzpicture}
\draw (0.5\textwidth, -3) node[text width = \textwidth] {\huge  \textsf{\textbf{Hydrodynamic theory of thermoelectric transport \\ \vspace{0.07in}  and negative magnetoresistance in Weyl semimetals}} };
\end{tikzpicture}
\end{equation*}
\begin{equation*}
\begin{tikzpicture}
\draw (0.5\textwidth, 0.1) node[text width=\textwidth] {\large \color{black} \textsf{Andrew Lucas,}$^{\color{\stylecolor} \mathsf{a}}$ \textsf{Richard A. Davison}$^{\color{\stylecolor} \mathsf{a}}$ \textsf{and Subir Sachdev}$^{\color{\stylecolor} \mathsf{a,b}}$};
\draw (0.5\textwidth, -0.5) node[text width=\textwidth] { $^{\color{\stylecolor} \mathsf{a}}$  \small\textsf{Department of Physics, Harvard University, Cambridge, MA 02138, USA}};
\draw (0.5\textwidth, -1) node[text width=\textwidth] { $^{\color{\stylecolor} \mathsf{a}}$  \small\textsf{Perimeter Institute for Theoretical Physics, Waterloo, Ontario N2L 2Y5, Canada}};
\end{tikzpicture}
\end{equation*}
\begin{equation*}
\begin{tikzpicture}
\draw (0, -13.1) node[right, text width=0.5\paperwidth] {\texttt{lucas@fas.harvard.edu} \\ \texttt{rdavison@physics.harvard.edu} \\ \texttt{sachdev@g.harvard.edu}  };
\draw (\textwidth, -13.1) node[left] {\textsf{\today}};
\end{tikzpicture}
\end{equation*}
\begin{equation*}
\begin{tikzpicture}
\draw[very thick, color=\stylecolor] (0.0\textwidth, -5.75) -- (0.99\textwidth, -5.75);
\draw (0.12\textwidth, -6.25) node[left] {\color{\stylecolor}  \textsf{\textbf{Abstract:}}};
\draw (0.53\textwidth, -6) node[below, text width=0.8\textwidth, text justified] {\small We present a theory of thermoelectric transport in weakly disordered Weyl semimetals where the electron-electron scattering time is faster than the electron-impurity scattering time. Our hydrodynamic theory consists of relativistic fluids at each Weyl node, coupled together by perturbatively small inter-valley scattering, and long-range Coulomb interactions. The conductivity matrix of our theory is Onsager reciprocal and positive-semidefinite. In addition to the usual axial anomaly, we account for the effects of a distinct, axial-gravitational anomaly expected to be present in Weyl semimetals.  Negative thermal magnetoresistance is a sharp, experimentally accessible signature of this axial-gravitational anomaly, even beyond the hydrodynamic limit.};
\end{tikzpicture}
\end{equation*}

\tableofcontents

\titleformat{\section}
  {\gdef\sectionlabel{}
   \Large\bfseries\scshape}
  {\gdef\sectionlabel{\thesection }}{0pt}
  {\begin{tikzpicture}[remember picture,overlay]
	\draw (1, 0) node[right] {\color{\stylecolor} \textsf{#1}};
	\fill[color=\stylecolor] (0,-0.35) rectangle (0.7, 0.35);
	\draw (0.35, 0) node {\color{white} \textsf{\sectionlabel}};
       \end{tikzpicture}
  }
\titlespacing*{\section}{0pt}{15pt}{15pt}

\begin{equation*}
\begin{tikzpicture}
\draw[very thick, color=\stylecolor] (0.0\textwidth, -5.75) -- (0.99\textwidth, -5.75);
\end{tikzpicture}
\end{equation*}

\section{Introduction}
The recent theoretical predictions \cite{vishwanath, bernevig1, bernevig2} and experimental discoveries \cite{soljacic, syxu, bqlv} of Weyl semimetals open up an exciting new solid state playground for exploring the physics of anomalous quantum field theories.    These anomalies can lead to very striking signatures in simple transport measurements.  Upon applying a magnetic field $\mathbf{B} = B\hat{\mathbf{z}}$ and measuring the electrical conductivity $\sigma_{zz}$ parallel to $\mathbf{B}$, one predicts $\sigma_{zz}$ has a contribution which grows as  $B^2$  \cite{nielsen, spivakson, burkov}.  This ``longitudinal negative magnetoresistance" is a direct signature of the anomaly associated with the Weyl points in momentum space. Similar results have also been predicted for thermal and thermoelectric transport \cite{fiete, spivak1510}.  Negative magnetoresistance in $\sigma$, with the predicted $B^2$ dependence, has been observed experimentally in many different materials \cite{hjkim, xiong, huangprx, czli, hli, czhang, hirschberger}.

So far, the theories of this negative magnetoresistance assume two facts about the dynamics of the quasiparticles of the Weyl semimetal.   Firstly, it is assumed that the quasiparticles are long lived, and that a kinetic description of their dynamics is valid.    Secondly, it is assumed that the dominant scattering mechanism is between quasiparticles and impurities or phonons.   In most simple crystals -- including Weyl semimetals -- it is likely that this description is reasonable.

However, there are exotic metals in which the quasiparticle-quasiparticle scattering time is much smaller than the quasiparticle-impurity/phonon scattering time.   In such a finite temperature metal, the complicated quantum dynamics of quasiparticles reduces to classical hydrodynamics of long lived quantities -- charge, energy and momentum -- on long time and length scales.   Most theoretical \cite{gurzhi1, spivak2006, andreev, tomadin, vignale, polini, levitovhydro} and experimental \cite{molenkamp, bandurin, mackenzie} work on such electron fluids studies the dynamics of (weakly interacting) Fermi liquids in ultrapure crystals.    As expected, the physics of a hydrodynamic electron fluid is qualitatively different from the kinetic regime where quasiparticle-impurity/phonon scattering dominates, and there are qualitatively distinct signatures to look for in experiments.

Experimental evidence for a strongly interacting quasirelativistic plasma of electrons and holes has recently emerged in graphene \cite{crossno,ghaharikim}.  The relativistic hydrodynamic theories necessary to understand this plasma are different from ordinary Fermi liquid theory \cite{hkms}, and lead to qualitatively different transport phenomena \cite{lucas3, lucasplasma}.  The hydrodynamics necessary to describe an electron fluid in a Weyl material, when the Fermi energy is close to a Weyl node, is similar to the hydrodynamics of the graphene plasma, though with additional effects related to anomalies \cite{surowka,oz}.  Such a quasirelativistic regime is where negative magnetoresistance is most pronounced \cite{burkov}, and also where interaction effects can be strongest, due to the lack of a large Fermi surface to provide effective screening.   

In this paper, we develop a minimal hydrodynamic model for direct current (dc) thermoelectric transport in a disordered, interacting Weyl semimetal, where the Fermi energy is close to the Weyl nodes.   The first hydrodynamic approach to transport in a Weyl semimetal may be found in \cite{landsteinerhydro} (see also \cite{Sadofyev:2010pr,Roychowdhury:2015jha}).   In contrast to these, our approach manifestly ensures that the conductivity matrix is positive-semidefinite and Onsager reciprocal.   We apply an infinitesimal electric field $E_i$ and temperature gradient $\partial_i T$ to a Weyl semimetal, and compute the total charge current $J_i$ and heat current $Q_i$ using hydrodynamics.  We then read off the thermoelectric conductivity matrix defined by \begin{equation}
\left(\begin{array}{c} J_i \\ Q_i \end{array}\right) = \left(\begin{array}{cc} \sigma_{ij} &\ \alpha_{ij}  \\ T\bar\alpha_{ij} &\ \bar\kappa_{ij} \end{array}\right)  \left(\begin{array}{c} E_j \\ -\partial_j T \end{array}\right).  \label{eq:thermoeldef}
\end{equation}
In the limit where disorder, magnetic field and intervalley scattering are perturbatively weak, we show that all conductivities may be written as the sum of a Drude conductivity for each valley fluid, and a correction due to intervalley scattering:  $e.g.~\sigma_{ij} = \sigma^{\mathrm{Drude}}_{ij} + \sigma^{\mathrm{anom}}_{ij}$. We present a general formula for the coefficient of $B^2$ in $\sigma^{\mathrm{anom}}_{zz}$:   the  quantitative dependence of this coefficient on temperature and electron density can be different from quasiparticle-based methods.    

While the qualitative form of our results (e.g.~$\sigma^{\mathrm{anom}}_{ij} \sim B_i B_j$) is very similar to that found using kinetic theory approaches \cite{spivakson, burkov, fiete, spivak1510}, we strongly emphasize that the physical interpretations are often quite different.   For example, the emergence of Drude conductivities in our model is not due to the existence of long lived quasiparticles, but due to the fact that momentum relaxation is a perturbatively slow process \cite{hkms, lucasMM}.    Furthermore, distinct anomalies are responsible for the negative magnetoresistance in electrical vs. thermal transport.   This remains true even beyond our strict hydrodynamic limit.

In this paper, we work in units where $\hbar=k_{\mathrm{B}}=e=1$.   We will also generally set the Fermi velocity  $v_{\mathrm{F}}=1$.   In our relativistic formalism,  the effective speed of light is set by $v_{\mathrm{F}}$.

\section{Weyl Hydrodynamics}

We begin by developing our hydrodynamic treatment of the electron fluid, assuming the chemical potential lies close to the charge neutrality point for every node.  For simplicity, we assume that the Weyl nodes are locally isotropic to reduce the number of effective parameters.  It is likely straightforward, though  tedious, to generalize and study anisotropic systems.

We will firstly review the hydrodynamic theory of a chiral fluid with an anomalous axial U(1) symmetry, derived in \cite{surowka, oz}.    Neglecting intervalley scattering, this theory describes the dynamics near one Weyl node.   The equations of relativistic chiral hydrodynamics are the conservation laws for charge, energy and momentum, modified by the external electromagnetic fields which we denote with $F_{\mu\nu}$.   On a curved space with Riemann tensor $R_{\alpha\beta\delta\gamma}$, they read \begin{subequations}\begin{align}
\nabla_\mu T^{\mu\nu} &= F^{\nu\mu}J_\mu - \frac{G}{16\mpi^2}\nabla_\mu \left[\varepsilon^{\rho\sigma\alpha\beta}F_{\rho\sigma}{R^{\nu\mu}}_{\alpha\beta}\right], \\
\nabla_\mu J^\mu &= -\frac{C}{8} \varepsilon_{\mu\nu\rho\sigma}F^{\mu\nu}F^{\rho\sigma} - \frac{G}{32\mpi^2}  \varepsilon^{\mu\nu\rho\sigma}{R^\alpha}_{\beta\mu\nu}{R^\beta}_{\alpha\rho\sigma},
\end{align}\end{subequations}
where $C$ is a coefficient related to the standard axial anomaly and $G$ is a coefficient related to an axial-gravitational anomaly \cite{jensen}.  For a Weyl fermion\begin{equation}\label{eq:kweyl}
C = \frac{k}{4\mpi^2},\;\;\;\;\; G = \frac{k}{24}, 
\end{equation} with $k\in\mathbb{Z}$ the Berry flux associated with the Weyl node \cite{landsteiner}.     $J^\mu$ and the energy-momentum tensor $T^{\mu\nu}$ are related to the hydrodynamic variables of chemical potential $\mu$, temperature $T$, and velocity $u^\mu$ in a tightly constrained way \cite{surowka, oz}, which we review in the SI.
We will take the background electromagnetic field to be \begin{equation}
F = B \mathrm{d}x \wedge \mathrm{d}y + \partial_i \mu_0 \mathrm{d}x_i \wedge \mathrm{d}t,
\end{equation}
with $B$ a constant.     Constant $B$ is required by Maxwell's equations for the external electromagnetic field in equilibrium, at leading order.   

A single chiral fluid cannot exist in a Weyl material.    Instead, enough Weyl nodes must exist so that the ``net'' $C$ for the material vanishes.  This follows mathematically from the fact that the Brillouin zone of a crystal is necessarily a compact manifold and so the sum of the Berry fluxes associated with each node must vanish -- this is the content of the Nielsen-Ninomiya theorem \cite{nielsen}.   Hence, we must consider the response of multiple chiral fluids when developing our theory of transport.


One might hope that so long as each chiral fluid has a well-behaved response,  then the net conductivities are simply additive.   This is not so:  the transport problem is ill-posed for a single chiral fluid, once we apply a background magnetic field.   To see this, suppose that we apply an electric field such that $\mathbf{E}\cdot\mathbf{B} \ne 0$.   Then, the total charge in the sample obeys \begin{equation}
\frac{\mathrm{d}Q_{\mathrm{tot}}}{\mathrm{d}t} = \int \mathrm{d}^3\mathbf{x}\; \partial_\mu J^\mu = C\mathbf{E}\cdot\mathbf{B} V_3,  \label{eq:NN}
\end{equation}
with $V_3$ the spatial volume of the metal.   Even at the linear response level, we see that there is a \emph{necessary} $\mathcal{O}(E)$ time-dependence to any solution to the hydrodynamic equations (with spatial directions periodically identified).  If there is no static solution to the equations of motion, then any dc conductivity is an ill-posed quantity to compute.    There is also energy production in a uniform temperature gradient, proportional to $G\nabla T \cdot \mathbf{B}$, even when $C=0$ (see the SI).

The physically relevant solution to this issue is that multiple Weyl nodes exist in a real material, and this means that we must consider the coupled response of multiple chiral fluids.  Rare intervalley processes mediated by phonons and/or impurities  couple these chiral fluids together \cite{spivakson} and make the transport problem far richer for Weyl fluids than for simpler quantum critical fluids, including the Dirac fluid \cite{lucas3}.   

We label each valley fluid quantity with the labels $ab\cdots$. For example, $u^\mu_a$ is the velocity of valley fluid $a$ .    To avoid being completely overwhelmed with free parameters, we only include coefficients at zeroth order in derivatives coupling distinct fluids together.   In fact, this will be sufficient to capture the negative magnetoresistance, as we explain in the next section.  Accounting for this coupling modifies the conservation equations to \begin{subequations}\label{eq:eommain}\begin{align}
\nabla_\mu J^\mu_a &= -\frac{C_a}{8} \varepsilon_{\mu\nu\rho\sigma}F^{\mu\nu}F^{\rho\sigma} -\frac{G_a}{32\mpi^2}  \varepsilon^{\mu\nu\rho\sigma}{R^\alpha}_{\beta\mu\nu}{R^\beta}_{\alpha\rho\sigma} - \sum_b\left[\mathcal{R}_{ab}\nu_b  + \mathcal{S}_{ab}\beta_b\right], \label{eq:eommain1}\\
\nabla_\mu T^{\mu \nu}_a &= F^{\nu\mu}J_{\mu a} - \frac{G_a}{16\mpi^2}\nabla_\mu \left[\varepsilon^{\rho\sigma\alpha\beta}F_{\rho\sigma}{R^{\nu\mu}}_{\alpha\beta}\right] + u^\nu_a \sum_b \left[\mathcal{U}_{ab}\nu_b  + \mathcal{V}_{ab}\beta_b\right],  \label{eq:eommain2}
\end{align}\end{subequations}
where we have defined $\beta_a \equiv 1/T_a$ and $\nu_a \equiv  \beta_a \mu_a$.
The transport problem is well-posed if \begin{equation}
\sum_a C_a = \sum_a G_a = 0.   \label{eq:sumc}
\end{equation}

The new coefficients $\mathcal{R},\mathcal{S},\mathcal{U}$ and $\mathcal{V}$ characterize the rate of the intervalley transfer of charge, energy and momentum due to relative imbalances in chemical potential or temperature.  In writing (\ref{eq:eommain}), we have chosen the intervalley scattering of energy and momentum to be relativistic. This makes the analysis easier as it preserves Lorentz covariance, but will not play an important role in our results. In particular, the intervalley momentum transfer processes are subleading effects in our theory of transport.

The gradient expansion may be different for each fluid, but we will assume that $J^\mu_a$ and $T^{\mu\nu}_a$ depend only on fluid $a$.   We require that \begin{equation}
\sum_{a\text{ or }b} \mathcal{R}_{ab} =\sum_{a\text{ or }b} \mathcal{S}_{ab} =\sum_{a\text{ or }b}\mathcal{U}_{ab} =\sum_{a\text{ or }b} \mathcal{V}_{ab} = 0.  \label{eq:markov}
\end{equation} This ensures that globally charge and energy are conserved, as well as that uniform shifts in the background chemical potential and/or temperature, for all fluids simultaneously, are exact zero modes of the equations of motion.

For simplicity in (\ref{eq:eommain}), we have implicitly assumed that the Weyl nodes are all at the same chemical potential in equilibrium.  This is generally not true for realistic Weyl materials.   As non-trivial issues in hydrodynamics already arise without making this generalization, we will stick to the case where all Weyl nodes are at the same chemical potential in equilibrium in this paper.

For the remainder of this paper, we will be interested in transport in flat spacetimes where $R_{\mu\nu\alpha\beta}=0$.   Except where otherwise stated, we will assume Minkowski space from now on.   Hence, for most purposes, we write partial derivatives $\partial_\mu$ rather than covariant derivatives $\nabla_\mu$. However, we will continue to use the covariant derivative $\nabla_\mu$ at intermediate steps of the calculations where it is necessary.

\subsection{Thermodynamic Constraints}
We will now derive the constraints on our hydrodynamic parameters which are imposed by demanding that the second law of thermodynamics is obeyed locally.   Without intervalley coupling processes, and at the ideal fluid level (derivative corrections, including $F_{\mu\nu}$, are neglected), the second law of thermodynamics implies that the total entropy current $s^\mu$ (where $s_a$ is the entropy density of fluid $a$) obeys (see e.g.~\cite{Jensen:2012jh,minwalla})
 \begin{equation}
\partial_\mu s^\mu = \partial_\mu \left( \sum_a s_a u_a^\mu \right) = 0.  \label{eq:2law1}
\end{equation}
In the more generic, non-ideal, case the right hand side of (\ref{eq:2law1}) must be non-negative.     In our theory of coupled chiral fluids, the right hand side of (\ref{eq:2law1}) does not vanish already at the ideal fluid level: \begin{equation}
\partial_\mu s^\mu = \sum_{ab} \left( \beta_a \left[\mathcal{U}_{ab}\nu_b + \mathcal{V}_{ab}\beta_b\right] + \nu_a \left[\mathcal{R}_{ab}\nu_b + \mathcal{S}_{ab}\beta_b\right]\right) \ge 0.  \label{eq:2law2}
\end{equation}
There is no possible change we can make to the entropy current that is local which can subtract off the right hand side of (\ref{eq:2law2}).   Hence, we demand that the matrix \begin{equation}
\mathsf{A} \equiv \left(\begin{array}{cc} \mathcal{R} &\ -\mathcal{S} \\ -\mathcal{U} &\ \mathcal{V} \end{array}\right),
\end{equation}
is positive semi-definite.   

Using standard arguments for Onsager reciprocity in statistical mechanics \cite{landaustat}, one can show that  $\mathsf{A} = \mathsf{A}^{\mathsf{T}}$.    In the SI, we will show using the memory matrix formalism \cite{forster, lucasMM} that whenever the quantum mechanical operators $n_a$ and $\epsilon_a$ are naturally defined:\begin{equation}
A_{IJ} = T \lim_{\omega\rightarrow 0} \frac{\mathrm{Im}\left(G^{\mathrm{R}}_{\dot{x}_I \dot{x}_J}(\omega)\right)}{\omega},
\end{equation}
where $x_I$ denotes ($n_a,\epsilon_a$) and dots denote time derivatives.  We also prove (\ref{eq:markov}), and the symmetry and positive-semidefiniteness of $\mathsf{A}$ through the memory matrix formalism, at the quantum mechanical level.

\subsection{Equilibrium Fluid Flow}
We now find an equilibrium solution to (\ref{eq:eommain}).   Beginning with the simple case of $B=0$, it is straightforward to see following \cite{lucas} that an equilibrium solution is \begin{subequations}\begin{align}
\mu_a &= \mu_0(\mathbf{x}),   \\
T_a &= T_0 = \text{constant}, \\
u_a^\mu &= (1,\mathbf{0}).  
\end{align}\end{subequations}
Indeed as pointed out in \cite{lucas, lucas3}, this exactly satisfies (\ref{eq:eommain}) neglecting the inter-valley and anomalous terms.     Using (\ref{eq:markov}) it is straightforward to see that the intervalley terms also vanish on this solution.   If $B=0$ then  $C\varepsilon_{\mu\nu\rho\sigma}F^{\mu\nu}F^{\rho\sigma}=0$, and hence this is an exact solution to the hydrodynamic equations.   We define the parameter $\xi$ as the typical correlation length of $\mu(\mathbf{x})$: roughly speaking, $\xi \sim |\mu_0|/|\partial_x \mu_0|$.

Following \cite{minwalla}, we can perturbatively construct a solution to the equations of motion when $B\ne 0$, assuming that \begin{equation}
B \ll T^2 \text{ and } 1 \ll \xi T.  \label{eq:small}
\end{equation}
Both of these assumptions are necessary for our hydrodynamic formalism to be physically sensible.  Using these assumptions, it is consistent at leading order to only change $v_{ia}\ne 0$, but to keep $\mu_a$ and $T_a$ the same: \begin{equation}
v_{za} = \frac{C_a\mu_0^2 B}{2(\epsilon_a+P_a)} + \frac{G_a T_0^2B}{\epsilon_a+P_a} \equiv \mathfrak{v}(\mu(x),T,B_i).  \label{eq:vza}
\end{equation}

It may seem surprising that in a single chiral fluid, there would be a non-vanishing charge current.   This is a well-known phenomenon called the chiral magnetic effect (for a recent review, see \cite{kharzeevreview}).   In our model, the net current flow is the sum of the valley contributions:
\begin{equation}
J^z= \sum_a J_a^z = \sum_a C_a \mu_0 B = 0,
\end{equation}
and so indeed, this complies with the expectation that the net current in a solid-state system will vanish in equilibrium, as discussed (in more generality) in \cite{landsteiner, vazifeh}.

\section{Thermoelectric Conductivity}
We now linearize the hydrodynamic equations around this equilibrium solution, applying infinitesimally small external electric fields $\tilde E_i$, and temperature gradients $\tilde \zeta_i  = -\partial_i \log T$
to the fluid.   Although we have placed an equals sign in this equation,  we stress that we will apply $\tilde \zeta_i$ in such a way we may apply a constant temperature gradient on a compact space (with periodic boundary conditions).   Applying a constant $\tilde E_i$ is simple, and corresponds to turning on an external electric field in $F_{\mu\nu}$.
Applying a constant $\tilde\zeta_i$ is more subtle, and can be done by changing the spacetime metric to \cite{hartnollads} \begin{equation}
\mathrm{d}s^2 = \eta_{\mu\nu} \mathrm{d}x^\mu \mathrm{d}x^\nu  - 2 \frac{\mathrm{e}^{-\mathrm{i}\omega t}}{-\mathrm{i}\omega}  \tilde\zeta_i \mathrm{d}x_i \mathrm{d}t.  \label{eq:zetametric}
\end{equation}
$\omega$ is a regulator, which we take to 0 at the end of the calculation.   This spacetime is flat ($R_{\alpha\beta\gamma\delta}=0$).  In order to account for both $\tilde E_i$ and $\tilde \zeta_i$,  the external gauge field is modified to $A + \tilde A$, where \begin{equation}
\tilde A_i = -\left(\tilde E_i - \mu_0(\mathbf{x}) \tilde\zeta_i\right)  \frac{\mathrm{e}^{-\mathrm{i}\omega t}}{-\mathrm{i}\omega}.  \label{eq:tildeA}
\end{equation}

The hydrodynamic equations (\ref{eq:eommain}) must then be solved in this modified background.  In linear response, the hydrodynamic variables become  \begin{subequations}\label{eq:31}\begin{align}
\mu_a &= \mu_0(\mathbf{x}) + \tilde \mu_a(\mathbf{x}), \\
T_a &= T_0 + \tilde T_a(\mathbf{x}), \\
u^\mu_a &\approx \left(1+v_{ia}\frac{\tilde\zeta_i\mathrm{e}^{-\mathrm{i}\omega t}}{\mathrm{i}\omega}, v_{ia} + \tilde v_{ia}\right).
\end{align}\end{subequations}
Note that tilded variables represent objects which are first order in linear response.    The correction to $u^t_a$ is necessary to ensure that $u^\mu u_\mu = -1$ is maintained.  In general we cannot solve the linearized equations analytically, except in the limit of perturbatively weak disorder and magnetic field strength.  

We assume that the inhomogeneity in the chemical potential is small:  \begin{equation}
 \mu_0 =  \bar\mu_0 + u  \hat \mu_0(\mathbf{x}),
\end{equation}
with $u\ll \bar\mu_0$ and  $T$.    $u$ is our perturbative parameter, and we assume that $\hat\mu_0$ is a zero-mean random function with unit variance.  We assume the scalings \begin{equation}
\label{eq:uscalings}
B\sim u^2 \;\; \text{ and } \;\; \mathcal{R},\mathcal{S},\mathcal{U},\mathcal{V}\sim u^{6}.
\end{equation}
The hydrodynamic equations can be solved perturbatively, and the charge and heat currents may be spatially averaged on this perturbative solution.   The computation is presented in the SI, and we present highlights here.   At leading order, the linearized hydrodynamic equations reduce to   
\begin{subequations}\label{eq:eomlinear2}\begin{align}
&\partial_i \left[n_a \tilde w_{ia} + \sigma_{\textsc{q}a}\left(\tilde E_i - \partial_i \tilde \mu_a - \frac{\mu_0}{T}(T\tilde\zeta_i - \partial_i\tilde T_a)\right)\right] = C_a \tilde E_i B_i -\sum_b \left[\mathcal{R}_{ab} \tilde \nu_b + \mathcal{S}_{ab} \tilde \beta_b\right], \\
&\partial_i \left[Ts_a \tilde w_{ia} -\mu_0 \sigma_{\textsc{q}a}\left(\tilde E_i - \partial_i \tilde \mu_a - \frac{\mu_0}{T}(T\tilde\zeta_i - \partial_i\tilde T_a)\right)\right]    =2G_a T_0^2 \tilde \zeta_i B_i  \notag\\
&+ \sum_b \left[(\mathcal{R}_{ab}\mu_0 + \mathcal{U}_{ab}) \tilde \nu_b + (\mathcal{S}_{ab}\mu_0 + \mathcal{V}_{ab})\tilde\beta_b\right], \\
&n_a(\partial_i \tilde \mu_a - \tilde E_i) + s_a(\partial_i \tilde T_a -  T\tilde \zeta_i)  = \varepsilon_{ijk} \tilde w_{ja} n_a B_k.  \label{eq:36c}
\end{align}\end{subequations}
We have defined \begin{equation}
\tilde v_{ia} = \tilde w_{ia} + \frac{\partial \mathfrak{v}_i}{\partial \mu} \tilde \mu_a +   \frac{\partial \mathfrak{v}_i}{\partial T} \tilde T_a.
\end{equation} 
$\tilde w_{ia}$ represents the fluid velocity after subtracting the contribution coming from (\ref{eq:vza}) in local thermal equilibrium.    

(\ref{eq:eomlinear2}) depends on $G_a$, despite the fact that our spacetime is flat.   This follows from the subtle fact that thermodynamic consistency of the anomalous quantum field theory on curved spacetimes requires that the axial-gravitational anomaly alters the thermodynamics of fluids on flat spacetimes \cite{jensen}.

The total charge current is $\tilde J^i = \sum_a \tilde J^i_{a}$,
and the total heat current is $\tilde Q^i = \sum_a \tilde T^{ti}_a - \mu_0 \tilde J^i$.
At leading order in perturbation theory, we find that the charge current in each valley fluid may be written as \begin{equation}
\tilde J^i_{a} = n_a \tilde{\mathbb{V}}_{ia} + C_a \tilde{\mathbb{M}}_{a} B_i,   \label{eq:Jtilde}
\end{equation}
and the heat current per valley,  $\tilde T^{ti}_a - \mu_0 \tilde J^i_a$, may be written as \begin{equation}
\tilde Q^i_a = Ts_a \tilde{\mathbb{V}}_{ia} + 2G_a T_0 \tilde{\mathbb{T}}_a B_i.   \label{eq:Qtilde}
\end{equation}
In the above expressions $\tilde{\mathbb{V}}_{ia}$ is a homogeneous $\mathcal{O}(u^{-2})$ contribution to $\tilde w_{ia}$, and $\tilde{\mathbb{M}}_a$ and $\tilde{\mathbb{T}}_a$ are $\mathcal{O}(u^{-4})$ homogeneous contributions to $\tilde \mu_a$ and $\tilde T_a$ respectively.

The thermoelectric conductivity matrix is:\begin{subequations}\label{eq:mainres}\begin{align}
\sigma_{xx} = \sigma_{yy} &= \sum_a \frac{n_a^2 \Gamma_a}{\Gamma_a^2 + B^2n_a^2}, \\
\sigma_{xy} &= \sum_a \frac{B n_a^3 }{\Gamma_a^2 + B^2n_a^2}, \\
\sigma_{zz} &= \sum_a \frac{n_a^2}{\Gamma_a} +   \mathfrak{s}B^2, \\
\bar\kappa_{xx} = \bar\kappa_{yy} &= \sum_a \frac{Ts_a^2 \Gamma_a}{\Gamma_a^2 + B^2n_a^2}, \\
\bar\kappa_{xy} &= \sum_a \frac{BT n_a s_a^2 }{\Gamma_a^2 + B^2n_a^2}, \\
\bar\kappa_{zz} &= \sum_a \frac{Ts_a^2}{\Gamma_a} +   \mathfrak{h}B^2, \\
\alpha_{xx} = \alpha_{yy} &= \sum_a \frac{n_a s_a \Gamma_a}{\Gamma_a^2 + B^2n_a^2}, \\
\alpha_{xy} = -\alpha_{yx} &= \sum_a \frac{B n_a^2s_a }{\Gamma_a^2 + B^2n_a^2}, \\
\alpha_{zz} &= \sum_a \frac{n_as_a}{\Gamma_a} +   \mathfrak{a}B^2,
\end{align}\end{subequations}
where we have defined the four parameters \begin{subequations}\label{eq:anomtimes}\begin{align}
\mathfrak{s} &\equiv T\left(\begin{array}{cc} C_a &\  C_a \mu  \end{array}\right) \left(\begin{array}{cc} \mathcal{R}_{ab} &\ -\mathcal{S}_{ab}  \\ -\mathcal{U}_{ab} &\ \mathcal{V}_{ab}\end{array}\right)^{-1} \left(\begin{array}{c} C_b \\ C_b \mu \end{array}\right), \\
\mathfrak{h} &\equiv 4T^4 \left(\begin{array}{cc} 0 &\  G_a   \end{array}\right) \left(\begin{array}{cc} \mathcal{R}_{ab} &\ -\mathcal{S}_{ab}  \\ -\mathcal{U}_{ab} &\ \mathcal{V}_{ab}\end{array}\right)^{-1} \left(\begin{array}{c} 0 \\ G_b  \end{array}\right), \\
\mathfrak{a} &\equiv 2T^2 \left(\begin{array}{cc} 0 &\  G_a  \end{array}\right) \left(\begin{array}{cc} \mathcal{R}_{ab} &\ -\mathcal{S}_{ab}  \\ -\mathcal{U}_{ab} &\ \mathcal{V}_{ab}\end{array}\right)^{-1} \left(\begin{array}{c} C_b \\ C_b \mu \end{array}\right),   \\
\Gamma_a &\equiv  \frac{T_0^2 \left(s_a (\partial n_a/\partial \mu) - n_a (\partial s_a/\partial \mu)\right)^2}{3\sigma_{\textsc{q}a}(\epsilon_a+P_a)^2} u^2.
\end{align}\end{subequations}
In these expressions, sums over valley indices are implicit.    Coefficients odd under $z\rightarrow -z$ (such as $\sigma_{xz}$) vanish. Note that all of the contributions to the conductivities listed above are of the same order $\mathcal{O}(u^{-2})$ in our perturbative expansion, explaining the particular scaling limit (\ref{eq:uscalings}) in $u$ that was taken.

We have not listed the full set of transport coefficients.   The unlisted transport coefficients are related to those in (\ref{eq:mainres}) by Onsager reciprocity:  \begin{subequations}\label{eq:onsager}\begin{align}
\sigma_{ij}(B) &= \sigma_{ji}(-B), \\
\bar\kappa_{ij}(B) &= \bar\kappa_{ji}(-B), \\
\alpha_{ij}(B) &= \bar\alpha_{ji}(-B).
\end{align}\end{subequations}
The symmetry of $\mathsf{A}$ is crucial in order for the final conductivity matrix to obey (\ref{eq:onsager}).

Evidently, the conductivities perpendicular to the magnetic field are Drude-like.  This follows from principles which are by now very well understood \cite{hkms, lucasMM}.   In these weakly disordered fluids,  the transport coefficients are only limited by the rate at which momentum relaxes due to the disordered chemical potential  $\Gamma_a/(\epsilon_a+P_a)$,  and/or the rate at which the magnetic field relaxes momentum (by ``rotating'' it in the $xy$ plane),  $Bn_a/(\epsilon_a+P_a)$. This latter energy scale is the hydrodynamic cyclotron frequency \cite{hkms}.   In our hydrodynamic theory, we can see this momentum ``bottleneck'' through the fact that the components of the charge and heat currents in (\ref{eq:Jtilde}) and (\ref{eq:Qtilde}), perpendicular to $B_i$, are proportional to the same fluid velocity $\tilde{\mathbb{V}}_{ia}\sim u^{-2}$ at leading order.   The transport of the fluid is dominated by the slow rate at which this large velocity can relax.   Since the heat current and charge current are proportional to this velocity field, the contribution of each valley fluid to $\sigma$, $\alpha$ and $\bar\kappa$ are all proportional to one another in the $xy$-plane.

The remaining non-vanishing transport coefficients are $\sigma_{zz}$, $\alpha_{zz}$ and $\bar\kappa_{zz}$.   From (\ref{eq:mainres}), we see that these conductivities are a sum of a Drude-like contribution (since this is transport parallel to the magnetic field, there is no magnetic momentum relaxation) from each valley, as before, and a new ``anomalous'' contribution which couples the valley fluids together.    This anomalous contribution has a qualitatively similar origin as that discovered in \cite{spivakson, landsteinerhydro}.    It can crudely be understood as follows:  the chemical potential and temperature imbalances $\tilde{\mathbb{M}}$ and $\tilde{\mathbb{T}}$ are proportional to $B$ and inversely proportional to $\mathsf{A}$, as the homogeneous contributions to the right hand side of (\ref{eq:eomlinear2}) cancel.    Such thermodynamic imbalances lead to corrections to valley fluid charge and heat currents, analogous to the chiral magnetic effect -- these are the linear in $B$ terms in (\ref{eq:Jtilde}) and (\ref{eq:Qtilde}).     Combining these scalings together immediately gives us the qualitative form of the anomalous contributions to the conductivity matrix.

The positive-semidefiniteness of the thermoelectric conductivity matrix is guaranteed.    Thinking of the conductivity matrices as a sum of the anomalous contribution and Drude contributions for each valley, it suffices to show each piece is positive-definite individually.   The Drude pieces are manifestly positive definite, as is well-known (it is an elementary exercise in linear algebra to confirm).   To show the anomalous pieces are positive-semidefinite, it suffices to show that 
$\mathfrak{s}\mathfrak{h} \ge T \mathfrak{a}^2$.  
This follows from (\ref{eq:anomtimes}), and the Cauchy-Schwarz inequality $(\mathbf{v}_1^{\mathsf{T}} \mathsf{A} \mathbf{v}_1)(\mathbf{v}_2^{\mathsf{T}} \mathsf{A} \mathbf{v}_2) \ge (\mathbf{v}_1^{\mathsf{T}} \mathsf{A} \mathbf{v}_2)^2$ for any vectors $\mathbf{v}_{1,2}$, and a symmetric, positive-semidefinite matrix $\mathsf{A}$.  These arguments also guarantee $\mathfrak{s},\mathfrak{h}>0$.

Our expression for the conductivity may seem ill-posed -- it explicitly depends on the matrix inverse $\mathsf{A}^{-1}$, but $\mathsf{A}$ is not invertible.    In fact, the kernel of $\mathsf{A}$ has two linearly independent vectors:  $(1_a, 1_a)$ and $(1_a,-1_a)$,   with $1_a$ denoting a vector with valley indices with each entry equal to 1.    However, in the final formula for the thermoelectric conductivities, which sums over all valley fluids, we see that $\mathsf{A}^{-1}$ is contracted with vectors which are orthogonal to the kernel of $\mathsf{A}$ due to (\ref{eq:sumc}).   The expression for the conductivities is therefore finite and unique.


We present a simple example of our theory for a fluid with two identical Weyl nodes of opposite chirality in the SI, along with a demonstration that the equations of motion are unchanged when we account for long-range Coulomb interactions, or impose electric fields and temperature gradients through boundary conditions in a finite domain.   Hence, the transport coefficients we have computed above are in fact those which will be measured in experiment.

In this paper, we used inhomogeneity in the chemical potential to relax momentum when $B=0$.  By following the hydrodynamic derivation in \cite{lucas}, other mechanisms for disorder likely lead to the same thermoelectric conductivities as reported in (\ref{eq:mainres}), but with a different formula for $\Gamma_a$.

\subsection{Violation of the Wiedemann-Franz Law}
The thermal conductivity $\kappa_{ij}$ usually measured in experiments is defined with the boundary conditions $\tilde J_i=0$ (as opposed to $\bar{\kappa}_{ij}$, which is defined with $\tilde E_i=0$).   This thermal conductivity is related to the elements of the transport matrix (\ref{eq:thermoeldef}) by \begin{equation}
\kappa_{ij} = \bar\kappa_{ij} - T\bar\alpha_{ik}\sigma^{-1}_{kl}\alpha_{lj}.
\end{equation} 
In an ordinary metal, the Wiedemann-Franz (WF) law states that \cite{ashcroft} \begin{equation}
\mathcal{L}_{ij} \equiv \frac{\kappa_{ij}}{T\sigma_{ij}} = \frac{\mpi^2}{3}.  \label{eq:WF}
\end{equation}
 The numerical constant of $\mpi^2/3$ comes from the assumption that the quasiparticles are fermions, and that the dominant interactions are between quasiparticles and phonons or impurities, but otherwise is robust to microscopic details.

In general, our model will violate the WF law.  Details of this computation are provided in the SI.   In general, the WF law is violated by an $\mathcal{O}(1)$ constant, which depends on the magnetic field $B$.   However, in the special case where we have valley fluids of opposite chirality but otherwise identical equations of state, we find that the transverse Lorenz ratios $\mathcal{L}_{xx}$, $\mathcal{L}_{xy}$, $\mathcal{L}_{yy}$ are all parametrically smaller than 1 (in fact, they vanish at leading order in perturbation theory).   In contrast, we find that $\mathcal{L}_{zz} \sim B^2$ at small $B$, and saturates to a finite number as $B$ becomes larger (but still $\ll T^2$).   This dramatic angular dependence of the WF law would be a sharp experimental test of our formalism in a strongly correlated Weyl material.

If the intervalley scattering rate is almost vanishing, the anomalous conductivities of a weakly interacting Weyl gas are still computable with our formalism.   Weak intravalley scattering processes  bring the ``Fermi liquid'' at each Weyl node to thermal equilibrium, and $\mathsf{A}$ may be computed via semiclassical kinetic theory.  Assuming that intervalley scattering occurs  elastically off of point-like impurities, we compute $\mathfrak{s}$, $\mathfrak{a}$ and $\mathfrak{h}$ in the SI.   We find that $\mathcal{L}^{\mathrm{anom}}_{zz} < \pi^2/3$, asymptotically approaching the WF law when $\mu \gg T$.  This is in contrast to the non-anomalous conductivities of a semiconductor, where under similar assumptions $\mathcal{L}_{zz}>\pi^2/3$ \cite{goldsmid}.   The Mott relation between $\alpha^{\mathrm{anom}}_{zz}$ and $\sigma^{\mathrm{anom}}_{zz}$ differs by an overall sign from the standard relation.    These discrepancies occur because we have assumed that elastic intervalley scattering is weaker than intravalley thermalization.   In contrast,  \cite{spivak1510} makes the opposite assumption when $\mu \gg T$, and so recovers all ordinary metallic phenomenology.   Even in this limit, negative thermal magnetoresistance is a consequence of non-vanishing $G_a$. 

\section{Outlook}
In this paper, we have systematically developed a hydrodynamic theory of thermoelectric transport in a Weyl semimetal where quasiparticle-quasiparticle scattering is faster than quasiparticle-impurity and/or quasiparticle-phonon scattering.    We have demonstrated the presence of longitudinal negative magnetoresistance in all thermoelectric conductivities.    New phenomenological parameters introduced in our classical model may be directly computed using the memory matrix formalism given a microscopic quantum mechanical model of a Weyl semimetal.   Our formalism is directly applicable to microscopic models of interacting Weyl semimetals where all relevant nodes are at the same Fermi energy.   Our model should be generalized to the case where different nodes are at different Fermi energies, though our main results about the nature of negative magnetoresistance likely do not change qualitatively.

Previously, exotic proposals have been put forth to measure the axial-gravitational anomaly in an experiment.   Measurements involving rotating cylinders of a Weyl semimetal have been proposed in \cite{landsteiner, landsteiner2},  and it is possible that the  rotational speed of neutron stars is related to this anomaly \cite{kaminski}.    A non-vanishing negative magnetoresistance in either $\alpha_{zz}$ or $\bar\kappa_{zz}$ is a direct experimental signature of the axial-gravitational anomaly.  It is exciting that a relatively mundane transport experiment on a Weyl semimetal is capable of detecting this novel anomaly for the first time.

\addcontentsline{toc}{section}{Acknowledgements}
\section*{Acknowledgements}
RAD~is supported by the Gordon and Betty Moore Foundation EPiQS Initiative through Grant GBMF\#4306.   AL and SS are supported by the NSF under Grant DMR-1360789 and MURI grant W911NF-14-1-0003 from ARO.   Research at Perimeter Institute is supported by the Government of Canada through Industry Canada and by the Province of Ontario through the Ministry of Research and Innovation. SS also acknowledges support from Cenovus Energy at Perimeter Institute.
\begin{appendix}

\section{Single Chiral Fluid}
To first order, the hydrodynamic gradient expansion of a chiral fluid reads \cite{surowka, oz} \begin{subequations}\label{eq:5b}\begin{align}
J^\mu &= nu^\mu - \sigma_{\textsc{q}} \mathcal{P}^{\mu\nu}\left(\nabla_\nu \mu - \frac{\mu}{T}\nabla_\nu T - F_{\nu\rho}u^\rho \right)  + \mathcal{D}_1\varepsilon^{\mu\nu\rho\sigma}u_\nu \nabla_\rho u_\sigma + \frac{\mathcal{D}_2}{2} \varepsilon^{\mu\nu\rho\sigma}u_\nu F_{\rho\sigma}, \\
T^{\mu\nu} &= (\epsilon+P)u^\mu u^\nu + P\eta^{\mu\nu} - \eta \mathcal{P}^{\mu\rho}\mathcal{P}^{\nu\sigma}(\nabla_\rho u_\sigma + \nabla_\sigma u_\rho) - \left(\zeta - \frac{2\eta}{3}\right) \mathcal{P}^{\mu\nu} \nabla_\rho u^\rho,  
\end{align}\end{subequations} where $\eta$ and $\zeta$ are shear and bulk viscosities, $\sigma_{\textsc{q}}$ is a ``quantum critical" conductivity,\begin{equation}
\mathcal{P}^{\mu\nu} = g^{\mu\nu} + u^\mu u^\nu, 
\end{equation}
and \begin{subequations}\begin{align}
\mathcal{D}_1 &= \frac{C\mu^2}{2} \left(1-\frac{2}{3}\frac{n\mu}{\epsilon+P}\right) - \frac{4G\mu nT^2}{\epsilon+P}, \\
\mathcal{D}_2 &= C\mu \left(1 - \frac{1}{2}\frac{n\mu}{\epsilon+P}\right) - \frac{GT^2n}{\epsilon+P}.
\end{align}\end{subequations}
There is a further coefficient that is allowed in $\mathcal{D}_1$ \cite{oz, jensen}, though it does not contribute to transport and so we will neglect it in this paper.   The entropy current is given by \begin{equation}
s^\mu \equiv (\epsilon+P)u^\mu - \frac{\mu}{T} J^\mu + \left(\frac{C\mu^3}{3T} + 2G\mu T\right)\varepsilon^{\mu\nu\rho\sigma}u_\nu \nabla_\rho u_\sigma   + \left( \frac{C\mu^2}{2T} + GT\right)  \frac{1}{2} \varepsilon^{\mu\nu\rho\sigma}u_\nu F_{\rho\sigma}.  \label{eq:entropycur}
\end{equation}

\section{Transport in the Weak Disorder Limit}
Here we present details of the computation of the thermoelectric conductivity matrix, using the notation for the perturbative transport computation presented in the main text.   At the first non-trivial order in an expansion at small $B$ and $1/\xi$ (assuming that they are of a similar magnitude),\footnote{It is important to only work to leading order in $1/\xi$ since we only know the background solution to leading order in $1/\xi$.}  the linearized hydrodynamic equations are \begin{subequations}\label{eq:eomlinear}\begin{align}
&\partial^i \left[n_a \tilde v_{ia} + \tilde n_a v_{ia}+ \sigma_{\textsc{q}a}\left(\tilde E_i + \varepsilon_{ijk}B^k\tilde v_{ja}- \partial_i \tilde \mu_a - \frac{\mu_a}{T}(T\tilde\zeta_i - \partial_i\tilde T_a)\right)\right]  - \partial_i \left\lbrace  \varepsilon^{ijk}\partial_j \mathcal{D}_{1a} \tilde v_{ka}  - B_i \tilde{\mathcal{D}}_{2a} \right. \notag \\
&\;\;\;\;\;\;\;\; \left. +\mathcal{D}_{2a} \varepsilon^{ijk} \tilde v_{ja} \partial_k \mu_0 \right\rbrace = C_a \tilde E_z  B  -\sum_b \left[\mathcal{R}_{ab} \tilde \nu_b + \mathcal{S}_{ab} \tilde \beta_b\right]  , \\
&\partial_i \left[Ts_a \tilde v_{ia} + (\tilde \epsilon_a + \tilde P_a - \mu_0 \tilde n_a )v_{ia} -\mu_0 \sigma_{\textsc{q}a}\left(\tilde E_i+ \varepsilon_{ijk}B^k\tilde v_{ja} - \partial_i \tilde \mu_a - \frac{\mu_a}{T}(T\tilde\zeta_i - \partial_i\tilde T_a)\right) - v_{ja} \eta_a \partial_j \tilde v_{ia} \right. \notag \\
&\;\;\;\;\;\;\;\; \left. - v_{ia}\left(\zeta_a+\frac{\eta_a}{3}\right)\partial_j \tilde v_{ja}\right] - \partial_i \left\lbrace\mu_0 B^i \tilde{\mathcal{D}}_{2a} - \mu_0 \mathcal{D}_{2a}\varepsilon^{ijk}\tilde v_{ja}\partial_k\mu_0 - \mu_0 \mathcal{D}_{1a}\varepsilon^{ijk}\partial_j \tilde v_{ka}\right\rbrace \notag \\
&\;\;\;\;\;\;\;\; = 2G_aT_0^2 B^i \tilde\zeta_i + \sum_b \left[(\mu_0 \mathcal{R}_{ab} + \mathcal{U}_{ab})\tilde\nu_b + (\mu_0 \mathcal{S}_{ab} + \mathcal{V}_{ab})\tilde \beta_b \right], \\
&n_a(\partial_i \tilde \mu_a - \tilde E_i) + s_a(\partial_i \tilde T_a - T\tilde \zeta_i) + \partial^j \left((\epsilon_a+P_a)(v_{ja}\tilde v_{ia} + v_{ia}\tilde v_{ja}) -\eta_a (\partial_j \tilde v_{ia} + \partial_i \tilde v_{ja}) - \left(\zeta_a-\frac{2\eta_a}{3}\right)\mdelta_{ij} \partial_k \tilde v_{ka}\right)  \notag \\
&\;\;\;\;\;\;  = \varepsilon_{ijk} \tilde J_a^j B^k  + v_{ia} \sum_b \left(\mathcal{U}_{ab} \tilde \nu_b + \mathcal{V}_{ab}\tilde \beta_b\right) \label{eq:eomlinear3} .
\end{align}\end{subequations}
These are respectively the equations of motion  for charge, heat and momentum.   In order to derive these equations, it is important to use covariant derivatives with respect to the metric.    We stress the importance of carefully deriving the $\tilde\zeta_i$-dependent terms in (\ref{eq:eomlinear}).   It is crucial that such terms be correctly accounted for in order for the resulting theory of transport to obey Onsager reciprocity at the perturbative level.

We have been able to remove many potential terms in the above equations which end up being proportional to $\varepsilon_{ijk} \partial_j \mu_0 \partial_k \mu_0 = \varepsilon_{ijk} \partial_j \mu_0 \partial_k v_l  =0$.    In (\ref{eq:eomlinear}), $\tilde n$ and other thermodynamic objects are to be interpreted as $\tilde n = (\partial_\mu n) \tilde \mu + (\partial_T n) \tilde T$, for example.  One finds terms $\sim \omega^{-1}$, which vanish identically assuming that the background is a solution to the hydrodynamic equations;  higher order terms in $\omega$ vanish upon taking $\omega \rightarrow 0$.   (\ref{eq:eomlinear}) is written in such a way that the terms on the left hand side are single-valley terms, with non-anomalous contributions to the charge and heat conservation laws written as the divergence of a current in square brackets, and anomalous contributions as the divergence of a current in curly brackets;  the terms on the right hand side of the charge and heat conservation laws are spatially homogeneous violations of the conservation laws.

The equations (\ref{eq:eomlinear}) are valid for a disordered chemical potential of any strength. We will now focus on the case where it is perturbatively small, and take the perturbative limit described in the main text.

   Let us split all of our perturbations into constants (zero modes of spatial momentum) and spatially fluctuating pieces: \begin{subequations}\begin{align}
\tilde v_{ia} &= \tilde w_{ia} + \frac{\partial \mathfrak{v}_i}{\partial \mu} \tilde \mu_a +   \frac{\partial \mathfrak{v}_i}{\partial T} \tilde T_a , \\
\tilde w_{ia} &= \tilde{\mathbb{V}}_{ia} + \tilde V_{ia}(\mathbf{x}), \\
\tilde \mu_a &= \tilde{\mathbb{M}}_a + \tilde M_a(\mathbf{x}) , \\
\tilde T_a &= \tilde{\mathbb{T}}_a + \tilde \theta_a(\mathbf{x}) .
\end{align}\end{subequations}
Recall that we defined $\mathfrak{v}$ to be the function of $\mu$ and $T$ which gives the equilibrium fluid velocity.  We will show self-consistently that the functions and constants introduced above  scale as \begin{equation}
\tilde{\mathbb{V}} \sim u^{-2}, \;\; \tilde{\mathbb{M}}\sim \tilde{\mathbb{T}} \sim u^{-4}, \;\;  \tilde{V}, \tilde{M}, \tilde{\theta} \sim u^{-1},   \label{eq:pertscale}
\end{equation}
at leading order in perturbation theory.   This will lead to charge and heat currents (and hence a conductivity matrix) which scale as $u^{-2}$.  To correctly capture the leading order response at small $u$, we do not need every term which has been retained in (\ref{eq:eomlinear}).   At leading order, linearized equations of motion reduce to those shown in the main text.
Upon replacing $\tilde v$ with $\tilde w$, the resulting equations have become much simpler.    Note that  terms proportional to $\varepsilon_{ijk}\mathfrak{v}_j B_k  =0$ (since $\mathfrak{v}_i \sim B_i$) can be dropped in the limit $B\rightarrow0$ and $\xi \rightarrow \infty$, as can viscous terms, which can be shown to contribute extra factors of $1/\xi$ to the answer \cite{lucas3}.

Next, we define the charge and heat currents in our hydrodynamic theory.   In an individual valley fluid, the leading order contributions to the charge current (\ref{eq:5b}) are \begin{equation}
\tilde J^i_{a} = n_a \tilde v_{ia} + \tilde n_a v_{ia} + \tilde{\mathcal{D}}_{2a} B_i  = n_a \tilde{\mathbb{V}}_{ia} + C_a \tilde{\mathbb{M}}_{a} B_i.   \label{eq:Jtilde}
\end{equation}
The last step follows from the definitions of $\mathfrak{v}$ and $\mathbb{V}$.   We assume that the total charge current is \begin{equation}
\tilde J^i = \sum_a \tilde J^i_{a}.   \label{eq:Jtildesum}
\end{equation}
The canonical definition of the global heat current for all valley fluids is \begin{equation}
\tilde Q^i = \tilde T^{ti} - \mu_0 \tilde J_i.
\end{equation}
In order to write $\tilde Q^i$ as a sum over valley contributions: \begin{equation}
\tilde Q^i = \sum_a \tilde Q^i_a,  \label{eq:Qtildesum}
\end{equation}
we define\begin{equation}
\tilde Q^i_a = \tilde T^{ti}_a - \mu_0 \tilde J^i_a.
\end{equation}
A simple computation reveals that at leading order in perturbation theory,\begin{equation}
\tilde Q^i_a = Ts_a \tilde{\mathbb{V}}_{ia} + 2G_a T_0 \tilde{\mathbb{T}}_a B_i.   \label{eq:Qtilde}
\end{equation}
$\tilde Q^i_a$  is not equivalent to the entropy current of an individual valley fluid, even at leading order.    

We now proceed to determine the spatially uniform responses $\tilde{\mathbb{V}}_{ia}$, $\tilde{\mathbb{M}}_a$ and $\tilde{\mathbb{T}}_a$ to leading order.  We begin by focusing on the inhomogeneous parts of the linearized equations.  It is simplest to do so in momentum space.  At the leading order $\mathcal{O}(u^{-1})$, the inhomogeneous equations of motion are \begin{subequations}\label{eq28}\begin{align}
\mathrm{i}k_i \left[n_a(\mathbf{k})\tilde{\mathbb{V}}_{ia} + n_a \tilde V_{ia}(\mathbf{k}) - \sigma_{\textsc{q}a} \mathrm{i}k_i \left(\tilde M_a(\mathbf{k}) - \frac{\mu_0}{T}\tilde \theta_a(\mathbf{k})\right)\right] &= 0, \\
\mathrm{i}k_i \left[Ts_a(\mathbf{k})\tilde{\mathbb{V}}_{ia}  + Ts_a \tilde V_{ia}(\mathbf{k})+ \mu_0 \sigma_{\textsc{q}a} \mathrm{i}k_i \left(\tilde M_a(\mathbf{k}) - \frac{\mu_0}{T}\tilde \theta_a(\mathbf{k})\right)\right] &= 0, \\
n_a \tilde M_a(\mathbf{k}) + s_a \tilde \theta_a(\mathbf{k})   &= 0.
\end{align}\end{subequations}
These equations are identical to those in \cite{lucas3} (with vanishing viscosity), but in one higher dimension.  Note that any term written without an explicit $\mathbf{k}$ dependence denotes the constant $\mathbf{k}=\mathbf{0}$ mode.   These equations give the following relations for the spatially dependent parts of the hydrodynamic variables \begin{subequations}\begin{align}
k_i \tilde V_{ia}(\mathbf{k}) &= -\frac{\mu_0 n_a(\mathbf{k}) + T_0 s_a(\mathbf{k})}{\epsilon_a+P_a} k_i \tilde{\mathbb{V}}_{ia}, \\
\tilde \theta_a(\mathbf{k}) &= \frac{\mathrm{i}k_i \tilde{\mathbb{V}}_{ia} T_0^2 n_a (s_a n_a(\mathbf{k}) - n_a s_a(\mathbf{k}))}{\sigma_{\textsc{q}a}k^2(\epsilon_a+P_a)^2}, \\
\tilde M_a(\mathbf{k}) &= -\frac{\mathrm{i}k_i \tilde{\mathbb{V}}_{ia} T_0^2 s_a (s_a n_a(\mathbf{k}) - n_a s_a(\mathbf{k}))}{\sigma_{\textsc{q}a}k^2(\epsilon_a+P_a)^2}.
\end{align}\end{subequations}
To determine the conductivities, we also require the leading order homogeneous components of the equations of motion.  Spatially integrating over the momentum conservation equation, we find the leading order equation (at order $\mathcal{O}(u^0)$)
\begin{equation}
\Gamma_{ija} \tilde{\mathbb{V}}_{ja} -n_a \tilde E_i  - s_aT_0 \tilde \zeta_i = \varepsilon_{ijk} n_a \tilde{\mathbb{V}}_{ja} B_k,  \label{eq:vtilde}
\end{equation}
with \begin{equation} \label{eq:gammahydro}
\Gamma_{ija} \equiv \sum_{\mathbf{k}} \frac{k_ik_j}{k^2} \frac{T_0^2 \left(s_a (\partial n_a/\partial \mu) - n_a (\partial s_a/\partial \mu) \right)^2}{\sigma_{\textsc{q}a}(\epsilon_a+P_a)^2} u^2 \left|\hat\mu(\mathbf{k})\right|^2.
\end{equation}
$\Gamma_{ija}$ is proportional to the rate at which momentum relaxes in the fluid due to the effects of the inhomogeneous chemical potential. Henceforth, we will assume isotropy for simplicity:  $\Gamma_{ija} \equiv \Gamma_a \mdelta_{ij}$.  It is manifest from the definition that $\Gamma_a>0$.  We can easily solve this equation for $\tilde{\mathbb{V}}_{ia}$.

Finally, to see the effects of the anomalies on hydrodynamic transport, we spatially average over the charge and heat conservation equations. At leading order $\mathcal{O}(u^2)$, this gives
\begin{subequations}\begin{align}
C_a BE_z &=  \sum_b \left[\mathcal{R}_{ab}  \tilde\nu_b + \mathcal{S}_{ab} \tilde\beta_b\right], \\
-C_a \mu_0 B E_z - 2G_a T_0^2 B \zeta_z  &=  \sum_b \left[ \mathcal{U}_{ab} \tilde\nu_b +  \mathcal{V}_{ab} \tilde\beta_b\right].
\end{align}\end{subequations}
Due to the anomalies, external temperature gradients and electric fields induce changes in the chemical potential and temperature of each fluid, which result in charge and heat flow. In the above equations, and for the rest of this paragraph, the fluctuations $\tilde \nu_b$ and $\tilde \beta_b$ denote the homogeneous parts of these objects, as it is only these which contribute at leading order to $\tilde{\mathbb{M}}$ and $\tilde{\mathbb{T}}$.  Hence, we find that \begin{equation}
\left(\begin{array}{c}  \tilde\nu_a \\ - \tilde\beta_a \end{array}\right) = \left(\begin{array}{cc}  \mathcal{R}_{ab} &\ -\mathcal{S}_{ab} \\ -\mathcal{U}_{ab} &\ \mathcal{V}_{ab} \end{array}\right)^{-1}  \left(\begin{array}{c}  C_b BE_z  \\  C_b \mu_0 B E_z + 2G_b T_0^2 B \tilde \zeta_z\end{array}\right).  \label{eq:numeans}
\end{equation}
We will find useful the relation \begin{equation}
\left(\begin{array}{c}  \tilde{\mathbb{M}} \\ \tilde{\mathbb{T}} \end{array}\right) = \left(\begin{array}{cc}  T &\ \mu T \\ 0 &\ T^2  \end{array}\right)  \left(\begin{array}{c}  \tilde\nu \\ - \tilde\beta \end{array}\right).  \label{eq:bMT}
\end{equation}

We are now ready to construct the thermoelectric conductivity matrix.   Combining the definition of the thermoelectric conductivity matrix,  (\ref{eq:Jtildesum}) and (\ref{eq:Qtildesum}) with our hydrodynamic results   (\ref{eq:Jtilde}), (\ref{eq:Qtilde}), (\ref{eq:vtilde}), (\ref{eq:numeans}) and (\ref{eq:bMT}) we obtain the thermoelectric conductivity matrix presented in the main text.

\section{Simple Example}
It is instructive to study the simplest possible system with an anomalous contribution to the conductivity.  This is a Weyl semimetal with 2 valley fluids, where the Berry flux \begin{equation}
k_1 = -k_2 = 1,
\end{equation}
and $C_{1,2}$ and $G_{1,2}$ are given by the results for a free Weyl fermion \cite{landsteiner}.  We also assume that the equations of state and disorder for each valley fluid are identical, so that $n_{1,2} = n$, $s_{1,2}=s$ and $\Gamma_{1,2}=\Gamma$.   Finally, we take the simplest possible ansatz for $\mathsf{A}$ consistent with symmetry, positive-definiteness and global conservation laws: \begin{subequations}\begin{align}
\mathcal{R} &= \left(\begin{array}{cc}  \mathcal{R}_0 &\ -\mathcal{R}_0 \\ -\mathcal{R}_0 &\ \mathcal{R}_0 \end{array}\right), \\
\mathcal{S} = \mathcal{U} &= \left(\begin{array}{cc}  \mathcal{S}_0 &\ -\mathcal{S}_0 \\ -\mathcal{S}_0 &\ \mathcal{S}_0 \end{array}\right), \\
\mathcal{V} &= \left(\begin{array}{cc}  \mathcal{V}_0 &\ -\mathcal{V}_0 \\ -\mathcal{V}_0 &\ \mathcal{V}_0 \end{array}\right),
\end{align}\end{subequations}
with positive-semidefiniteness of $\mathsf{A}$ imposing $\mathcal{R}_0,\mathcal{V}_0\ge 0$ and \begin{equation}
\mathcal{R}_0\mathcal{V}_0\ge \mathcal{S}_0^2. 
\end{equation}
We find the thermoelectric conductivities \begin{subequations}\begin{align}
\sigma_{xx} = \sigma_{yy} &= 2\frac{n^2 \Gamma}{\Gamma^2 + B^2n^2}, \\
\sigma_{xy} &= 2 \frac{B n^3 }{\Gamma^2 + B^2n^2}, \\
\sigma_{xz}=\sigma_{yz} &=0, \\
\sigma_{zz} &= 2 \frac{n^2}{\Gamma} +   \frac{TB^2(\mathcal{R}_0\mu^2 + 2\mathcal{S}_0\mu + \mathcal{V}_0)}{16\mpi^4(\mathcal{R}_0\mathcal{V}_0-\mathcal{S}_0^2)}, \\
\bar\kappa_{xx} = \bar\kappa_{yy} &= 2\frac{Ts^2 \Gamma}{\Gamma^2 + B^2n^2}, \\
\bar\kappa_{xy} &= 2\frac{BT n s^2 }{\Gamma^2 + B^2n^2}, \\
\bar\kappa_{xz}=\bar\kappa_{yz} &=0, \\
\bar\kappa_{zz} &= 2\frac{Ts^2}{\Gamma} +   \frac{T^4B^2 \mathcal{R}_0}{144(\mathcal{R}_0\mathcal{V}_0-\mathcal{S}_0^2)}, \\
\alpha_{xx} = \alpha_{yy} &= 2\frac{n s \Gamma}{\Gamma^2 + B^2n^2}, \\
\alpha_{xy}= -\alpha_{yx} &=2\frac{B n^2s }{\Gamma^2 + B^2n^2}, \\
\alpha_{xz}=\alpha_{yz} &=0, \\
\alpha_{zz} &= 2 \frac{ns}{\Gamma} +  \frac{T^2B^2(\mathcal{R}_0\mu + \mathcal{S}_0)}{48\mpi^2 (\mathcal{R}_0\mathcal{V}_0-\mathcal{S}_0^2)}.
\end{align}\end{subequations}
As expected due to the matrix inverse in the expressions for $\mathfrak{s}$, $\mathfrak{a}$ and $\mathfrak{h}$, we see that the anomalous contributions to the conductivities depend on the intervalley scattering rates for charge and energy in a rather complicated way.

\section{Imposing External Sources Through Boundary Conditions}
The derivation of the thermoelectric conductivity matrix presented above applied $\tilde E_i$ and $\tilde \zeta_i$  by particular deformations to background fields.   As in \cite{lucas}, one might also wish to impose electric fields and temperature gradients in a space with boundaries, as is done in a real experiment.  In this case, we do not need to deform the metric from Minkowski space,  nor the external gauge field, as we did in the main text.  

For example, let us keep the $x$ and $y$ directions periodic, but consider a Weyl fluid in the domain $0\le z \le L$, subject to the boundary conditions \begin{subequations}\begin{align}
\mu (z=0) &= \mu_0, \;\;\; \mu(z=L) = \mu_0 -\tilde E_z L, \\
T(z=0) &= T_0, \;\;\;  T(z=L) =T_0 -\tilde \zeta_z T_0 L.
\end{align}\end{subequations}
The hydrodynamic variables become \begin{subequations}\label{eq:31new}\begin{align}
\mu_a &=  \mu_0 + \tilde \mu_a - \tilde E_z z, \\
T_a &= T_0 + \tilde T_a - T_0 \tilde \zeta_zz, \\
u^\mu_a &= (1, v_{ia} + \tilde v_{ia}),
\end{align}\end{subequations}
and, in linear response, we can also arrive at (\ref{eq:eomlinear}).   The simplest way to see this as follows.  In equilibrium the hydrodynamic equations are satisfied (at leading order in $B$ and $\xi$).   After taking spatial derivatives in $\partial_\mu J^\mu_a$ (for example), it is possible to obtain terms of the form $-\tilde E_z z \times \partial_i \mu_0$ which are linear in $z$.    However, all such terms must identically cancel, because the background solution is independent of a global spatial shift in $\mu$ and $T$.   We have, in fact, already seen this explicitly -- the coefficients of $\mathbb{M}_a$ and $\mathbb{T}_a$ in the charge and heat currents (\ref{eq:Jtilde}) and (\ref{eq:Qtilde}) are all independent of $\mathbf{x}$.

Hence, upon plugging in (\ref{eq:31new}) into the equations of motion,  the only terms which do not vanish at leading order in $B$ and $1/\xi$ are \begin{subequations}\label{eq:eomlinearbc}\begin{align}
\partial_z J^z_a &= \partial_z \left(-C_a B E_z z + \cdots\right) = -\sum_b \left[\mathcal{R}_{ab} \tilde \nu_b + \mathcal{S}_{ab} \tilde \beta_b\right]  \\
\partial_z \left(T^{tz}_a - \mu_0 J^z_a\right) &= \partial_z \left(- 2G_a B T_0^2 \tilde\zeta_z z+\cdots\right) = \sum_b \left[(\mu_0 \mathcal{R}_{ab} + \mathcal{U}_{ab})\tilde\nu_b + (\mu_0 \mathcal{S}_{ab} + \mathcal{V}_{ab})\tilde \beta_b \right], \\
\partial_z T^{zz}_a &=  -n_a\tilde E_z - T_0 s_a\tilde\zeta_z + \cdots
\end{align}\end{subequations}
The $\cdots$ terms above are linear in $\tilde \mu_a$, $\tilde T_a$ or $\tilde v_{ia}$, and are the same as found in (\ref{eq:eomlinear}).   Upon comparing with (\ref{eq:eomlinear}), we see that the source ($\tilde E_z$ and $\tilde \zeta_z$) terms are identical.

Hence our equations of motion (\ref{eq:eomlinear}) are unchanged, and our perturbative theory of transport  can be recovered regardless of the choice of boundary conditions.   This is important as experiments will always impose temperature gradients through the boundary conditions on a finite domain.

\section{Coulomb Screening}\label{sec:coulomb}

Coulomb screening alters the electric field seen by the charges. In our equilibrium solution, it leads to an effective change in $\mu_0(\mathbf{x})$, the disorder profile seen by the fluid. We may account for it by replacing
\begin{equation}
\mu_0 \rightarrow  \mu_0 - \varphi \equiv \mu_0  - \int \mathrm{d}^3 \mathbf{y} \; K(\mathbf{x};\mathbf{y}) \sum_a n_a(\mathbf{y}),  \label{eq:coulomb}
\end{equation}
with $K \sim 1/r$ the Coulomb kernel (its precise form is not important, and we could include thermal screening effects if we wish). However, as pointed out in \cite{lucas3, lucasplasma}, by simply redefining $\mu_0$ to be the equilibrium electrochemical potential, one can neglect this effect.

We must still account for the effects of Coulomb screening on the linear response around the equilibrium state.   In our perturbative formalism, the leading order conductivities are governed by the equations of motion (\ref{eq:eomlinear}), and the simplification in the main text.     We may account for Coulomb screening in these equations by modifying the external electric field to \begin{equation}
\tilde E_i \rightarrow \tilde E_i -  \partial_i \tilde \varphi,
\end{equation}
where $\tilde \varphi$ is the convolution of the Coulomb kernel with $\sum_a \tilde n_a$.    The equations of motion then become \begin{subequations}\label{eq:coueom}\begin{align} 
\partial_i \left[n_a \tilde w_{ia} + \sigma_{\textsc{q}a}\left(\tilde E_i - \partial_i \tilde \Phi_a - \frac{\mu_0}{T}(T\tilde\zeta_i - \partial_i\tilde T_a)\right)\right] &= C_a \tilde E_i B_i -\sum_b \left[\mathcal{R}_{ab} \tilde \nu_b + \mathcal{S}_{ab} \tilde \beta_b\right], \\
\partial_i \left[Ts_a \tilde w_{ia} -\mu_0 \sigma_{\textsc{q}a}\left(\tilde E_i - \partial_i \tilde \Phi_a - \frac{\mu_0}{T}(T\tilde\zeta_i - \partial_i\tilde T_a)\right)\right] &= 2G_a T_0^2 \tilde \zeta_i B_i   \notag \\
&+ \sum_b \left[(\mathcal{R}_{ab}\mu_0 + \mathcal{U}_{ab}) \tilde \nu_b + (\mathcal{S}_{ab}\mu_0 + \mathcal{V}_{ab})\tilde\beta_b\right], \\
n_a(\partial_i \tilde \Phi_a - \tilde E_i) + s_a(\partial_i \tilde T_a -  T\tilde \zeta_i)  &= \varepsilon_{ijk} \tilde w_{ja} n_a B_k, 
\end{align}\end{subequations}
where we have defined \begin{equation}
\tilde \Phi_a \equiv \tilde \varphi + \tilde \mu_a.
\end{equation}
We have neglected the contribution of the Coulomb kernel to the anomalous creation of charge in a single valley in (\ref{eq:coueom}).   This is because, in our perturbative limit, only the homogeneous part of this term is important.   Since \begin{equation}
\tilde \nu_a = \frac{\tilde \mu_a}{T_0} + \tilde \beta_a \mu_0 = \frac{\tilde \Phi_a - \tilde \varphi}{T_0} + \tilde \beta_a \mu_0, 
\end{equation}
it follows from the fact that $\sum_b \mathcal{R}_{ab}=\sum_b \mathcal{S}_{ab}=0$  that the $\tilde\varphi$-dependent corrections to the inter-valley terms exactly cancel.   Mathematically, we now see that (\ref{eq:coueom}) are the same as the linearized equations of motion in the main text, up to a relabeling of the variables.    The long-range Coulomb interactions, introduced in hydrodynamics through $F_{\mu\nu}$,  do not alter our definitions of the charge current (\ref{eq:Jtilde}) or the heat current (\ref{eq:Qtilde}) at leading order in perturbation theory.  Hence, our expressions for the conductivities are not affected by long-range Coulomb interactions, confirming our claim in the main text. The interactions may alter the specific values of the parameters in the hydrodynamic equations (\ref{eq:5b}).\footnote{More carefully, if we place our equations on a periodic space, where the transport problem is still well-posed,  then the boundary conditions on $\tilde v_{ia}$, $\tilde \mu_a$ and $\tilde T_a$ are all periodic boundary conditions.   Hence, $\tilde w_{ia}$, $\tilde\Phi_a$ and $\tilde T_a$ all have periodic boundary conditions and so the change of variables between the linearized equations presented in the main text and (\ref{eq:coueom}) does not affect the transport problem even via non-trivial boundary conditions.}

Finite frequency transport is generally sensitive to long-range Coulomb interactions, although in (disordered) charge-neutral systems the effect is likely much more suppressed (see \cite{lucasplasma} for a recent discussion in two spatial dimensions).

\section{Violation of the Wiedemann-Franz Law in the Hydrodynamic Regime}

Since $\sigma_{ij}$, $\alpha_{ij}$  and $\bar\kappa_{ij}$ are block diagonal in our perturbative hydrodynamic formalism, $\kappa_{ij}$ will be as well.   We begin by focusing on the longitudinal ($zz$) conductivities.    A simple computation gives \begin{equation}
\kappa_{zz} = T\sum_a \frac{s_a^2}{\Gamma_a} + \mathfrak{h}B^2 - T\left(\mathfrak{a}B^2 + \sum_a \frac{s_an_a}{\Gamma_a}\right)^2 \left(\mathfrak{s}B^2 + \sum_a \frac{n_a^2}{\Gamma_a}\right)^{-1}.
\end{equation}
Firstly, consider the case $B=0$.  In this case, there are two possibilities of interest.   If\footnote{This is a stronger statement than necessary for this equation to hold for $\kappa_{zz}$.   It is sufficient for the ratio $s_a/n_a$ to be identical for all valley fluids for (\ref{eq:kappa0}) to hold.   However, this stricter requirement is necessary for (\ref{eq:kappa0B}) to hold.}   \begin{equation}
s_a= s\;\;\;\; \text{ and } \;\;\;\; n_a=n, \label{eq:indis}
\end{equation}
 for all valley fluids,  then \begin{equation}
 \kappa_{zz}(B=0) \sim \mathcal{O}\left(u^0\right), \label{eq:kappa0}
\end{equation}
is subleading in perturbation theory.   Hence, assuming $n\ne 0$ (i.e., the system is at finite charge density), we find that $\mathcal{L}_{zz} \ll \mathcal{L}_{\mathrm{WF}}$ in the perturbative limit $u\rightarrow 0$.   That a charged fluid has a highly suppressed $\kappa$ is by now a well-appreciated effect in normal relativistic fluids with a single valley \cite{hkms, maissam}.  If the valley fluids are indistinguishable as in (\ref{eq:indis}), then they behave as a ``single valley" at $B=0$ and so the considerations of \cite{hkms, maissam} apply here.   The reason that (\ref{eq:kappa0}) is small relative to $\bar\kappa_{zz}$ is that the boundary condition $\tilde J=0$ forces us to set (at leading order in $u$) the velocity $\tilde{\mathbb{V}}=0$, which means that both the leading order charge and heat currents vanish.

However, at a non-zero value of B, the leading order contribution does not vanish: $\kappa_{zz}(B) \sim u^{-2}$.   In particular, as $B\rightarrow 0$ \begin{equation}
\kappa_{zz}(B\rightarrow0) \approx \left(\mathfrak{h} + Ts\frac{s\mathfrak{s} - 2n\mathfrak{a}}{n^2}\right) B^2,
\end{equation} 
while at larger $B$ (such that $n^2/\mathfrak{s},\;sn/\mathfrak{a},\;Ts^2/\mathfrak{h}\ll \Gamma B^2$, while keeping $B\ll T^2$) 
\begin{equation}
\kappa_{zz} \approx \left(\mathfrak{h} - \frac{T\mathfrak{a}^2}{\mathfrak{s}}\right)B^2.
\end{equation}
Hence, as $B\rightarrow 0$, $\mathcal{L}_{zz} \sim B^2$ is parametrically small, but at larger $B$ it approaches a finite value \begin{equation}
\mathcal{L}_{zz} \rightarrow \frac{\mathfrak{h}}{T\mathfrak{s}} - \frac{\mathfrak{a}^2}{\mathfrak{s}^2}. \label{eq:LzzB1}
\end{equation}

If (\ref{eq:indis}) does not hold, then we instead find that $\kappa_{zz}(B)$ is finite and $\mathcal{O}(u^{-2})$, just as $\sigma_{zz}(B)$.  Hence, we find that the Lorenz ratio $\mathcal{L}_{zz}$ is generally $\mathcal{O}(1)$ and there are no parametric violations.   However, the Wiedemann-Franz law will not hold in any quantitative sense, and $\mathcal{L}_{zz}$ can easily be $B$-dependent.

In the $xy$-plane, the Wiedemann-Franz law has a somewhat similar fate.   If (\ref{eq:indis}) holds, then we find that the leading order contributions to the thermal conductivities all vanish at leading order, so that \begin{equation}
\kappa_{xx}(B) \sim \kappa_{yy}(B) \sim \kappa_{xy}(B) \sim \kappa_{yx}(B) \sim \mathcal{O}(u^0),  \label{eq:kappa0B}
\end{equation}
at all values of $B$, and so the corresponding components of $\kappa_{ij}$ will be parametrically small.  If (\ref{eq:indis}) does not hold, then $\kappa_{ij}(B) \sim u^{-2}$ is never parametrically small, and so the Wiedemann-Franz law will not be violated parametrically, but will be violated by an $\mathcal{O}(1)$ $B$-dependent function.

\section{Weak Intervalley Scattering in a Weakly Interacting Weyl Gas}
As noted in the main text, it is possible to employ our hydrodynamic formalism even when the fluids at each node are weakly interacting.   In fact, the only requirement to use our formalism for computing the anomalous thermoelectric conductivities is that the time scales set by $\mathsf{A}$ are the longest time scales in the problem (in particular: slower than any thermalization time scale within a given valley fluid).    Now, we consider a weakly interacting Weyl semimetal with long lived quasiparticles, but where the intervalley scattering is weak enough that our formalism nevertheless is valid.     We will employ semiclassical kinetic theory to find relations between $\mathcal{R}_0$, $\mathcal{S}_0$ and $\mathcal{V}_0$ under reasonable assumptions.

For simplicity, as in our simple example above, we will consider a pair of nodes with opposite Berry flux, but otherwise identical equations of state.     We suppose that the two Weyl nodes (located at the same Fermi energy) are at points $\mathbf{K}_{1,2}$ in the Brillouin zone, with $|\mathbf{K}_1-\mathbf{K}_2| \gg \mu, T$.    

Denote with $f(\mathbf{k})$ the number density of quasiparticles at momentum $\mathbf{k}$.   Under very basic assumptions about the nature of weak scattering off of impurities, assuming all scattering off of impurities is elastic, one finds the kinetic theory result \cite{ashcroft} \begin{equation}
\frac{\mathrm{d}f(\mathbf{k})}{\mathrm{d}t} = \int \frac{\mathrm{d}^3\mathbf{k}^\prime }{(2\mpi)^3} W(\mathbf{k},\mathbf{k^\prime}) \left[f(\mathbf{k}^\prime) - f(\mathbf{k})\right]
\end{equation}
where for simplicity we assume spatial homogeneity.  $W(\mathbf{k},\mathbf{k}^\prime)$ denotes the scattering rate of a quasiparticle from momentum $\mathbf{k}$ to $\mathbf{k}^\prime$ -- under the assumptions listed above, this is a symmetric function which may be perturbatively computed using Fermi's golden rule \cite{ashcroft}.   Since scattering is elastic, we have $W(\mathbf{k},\mathbf{q}) \sim \mdelta(k-q)$.   Using that\footnote{In these equations we have noted that the integrand is odd under exchanging $\mathbf{k}$ and $\mathbf{q}$ (and thus vanishes upon integration over $\mathbf{k}$ and $\mathbf{q}$) if $\mathbf{k}$ and $\mathbf{q}$ belong to the same node.     } \begin{subequations}\begin{align}
\frac{\mathrm{d}n_1}{\mathrm{d}t} = -\frac{\mathrm{d}n_2}{\mathrm{d}t} &= \int \left.\frac{\mathrm{d}^3\mathbf{k}}{(2\mpi)^3}\right|_{\text{node 1}}\left.\frac{\mathrm{d}^3\mathbf{q}}{(2\mpi)^3}\right|_{\text{node 2}} W(\mathbf{k},\mathbf{q}) \left[f(\mathbf{q}) - f(\mathbf{k})\right] \\
\frac{\mathrm{d}\epsilon_1}{\mathrm{d}t} = -\frac{\mathrm{d}\epsilon_2}{\mathrm{d}t}&= \int \left.\frac{\mathrm{d}^3\mathbf{k}}{(2\mpi)^3}\right|_{\text{node 1}}\left.\frac{\mathrm{d}^3\mathbf{q}}{(2\mpi)^3}\right|_{\text{node 2}} W(\mathbf{k},\mathbf{q}) \left[f(\mathbf{q}) - f(\mathbf{k})\right] |\mathbf{k}|
\end{align}\end{subequations}
In the above integrals, the subscript node 1 implies that the momentum integral is shifted so that $\mathbf{k} = \mathbf{0}$ at the point $\mathbf{K}_1$;  a similar statement holds for node 2.   All low energy quasiparticles are readily identified as either belonging to node 1 or 2.   Since we have set $v_{\mathrm{F}}=1$, the energy of a quasiparticle of momentum $\mathbf{k}$ (near node 1) is simply $|\mathbf{k}|$;  a similar statement holds for quasiparticles near node 2.

The simplest possible assumption is that \begin{equation}
W(\mathbf{k},\mathbf{q}) = W_0(k) \mdelta(|\mathbf{k}|-|\mathbf{q}|),
\end{equation}
and we may further take $W_0$ to be a constant if we desire.   For our hydrodynamic description to be valid, nodes 1 and 2 are in thermal equilibrium, up to a relative infinitesimal shift in temperature and chemical potential.   For simplicity suppose that node 1 is at a different $\beta$ and $\nu$.   Then the infinitesimal change in the rate of charge and energy transfer is \begin{subequations}\begin{align}
\frac{\mathrm{d}n_1}{\mathrm{d}t}  &= -\int \left.\frac{\mathrm{d}^3\mathbf{k}}{(2\mpi)^3}\right|_{\text{node 1}}\left.\frac{\mathrm{d}^3\mathbf{q}}{(2\mpi)^3}\right|_{\text{node 2}} W(\mathbf{k},\mathbf{q}) n_{\mathrm{F}}(\beta k - \nu) \\
\frac{\mathrm{d}\epsilon_1}{\mathrm{d}t}  &= -\int \left.\frac{\mathrm{d}^3\mathbf{k}}{(2\mpi)^3}\right|_{\text{node 1}}\left.\frac{\mathrm{d}^3\mathbf{q}}{(2\mpi)^3}\right|_{\text{node 2}} W(\mathbf{k},\mathbf{q}) n_{\mathrm{F}}(\beta k - \nu)k
\end{align}\end{subequations}
We can now read off  \begin{subequations}\begin{align}
\mathcal{R}_0 &= \int \left.\frac{\mathrm{d}^3\mathbf{k}}{(2\mpi)^3}\right|_{\text{node 1}}\left.\frac{\mathrm{d}^3\mathbf{q}}{(2\mpi)^3}\right|_{\text{node 2}} W(\mathbf{k},\mathbf{q}) (-n_{\mathrm{F}}^\prime (\beta k - \nu)), \\
\mathcal{S}_0 = \mathcal{U}_0 &= -\int \left.\frac{\mathrm{d}^3\mathbf{k}}{(2\mpi)^3}\right|_{\text{node 1}}\left.\frac{\mathrm{d}^3\mathbf{q}}{(2\mpi)^3}\right|_{\text{node 2}} W(\mathbf{k},\mathbf{q}) (-n_{\mathrm{F}}^\prime (\beta k - \nu)) k, \\
\mathcal{V}_0 &= \int \left.\frac{\mathrm{d}^3\mathbf{k}}{(2\mpi)^3}\right|_{\text{node 1}}\left.\frac{\mathrm{d}^3\mathbf{q}}{(2\mpi)^3}\right|_{\text{node 2}} W(\mathbf{k},\mathbf{q}) (-n_{\mathrm{F}}^\prime (\beta k - \nu))k^2.
\end{align}\end{subequations}
Our kinetic theory computation gives $\mathcal{S}_0 = \mathcal{U}_0$, as required by general quantum mechanical principles, and serves as a consistency check on our kinetic theory approximations.   

As is reasonable for many materials, we first approximate that  $\mu \gg T$.  Using the Sommerfeld expansion of the Fermi function, we find that the leading and next-to-leading order terms as $T/\mu \rightarrow 0$ are: \begin{subequations}\begin{align}
\mathcal{R}_0 &\approx T A(\mu) + \frac{A^{\prime\prime}(\mu)}{2}\frac{\mpi^2}{3}T^3, \\
\mathcal{S}_0 &\approx -T\mu A(\mu)  - \mu \frac{A^{\prime\prime}(\mu)}{2}\frac{\mpi^2}{3}T^3 - A^{\prime}(\mu)\frac{\mpi^2}{3}T^3  \\
\mathcal{V}_0 &\approx T \mu^2 A(\mu)  + \frac{\mpi^2}{3}T^3\left(A(\mu) + 2\mu A^\prime(\mu) + \frac{A^{\prime\prime}(\mu)}{2}\mu^2\right)
\end{align}\end{subequations}
where we have defined \begin{equation}
A(\mu) = \frac{\mu^4}{4\mpi^4} W_0(\mu).
\end{equation}
The leading order anomalous conductivities are: \begin{subequations}\begin{align}
\sigma_{zz} & = \frac{B^2}{16\mpi^4} \frac{1}{A(\mu)}, \\
\alpha_{zz} &=  \frac{\mpi^2 T}{3} \frac{\partial \sigma_{zz}(\mu,T=0)}{\partial \mu} , \\
\bar\kappa_{zz} &= \frac{\mpi^2 T}{3} \sigma_{zz}.
\end{align}
\end{subequations}

Remarkably, we recover the Wiedemann-Franz law exactly in the limit $\mu \gg T$.   This is rather surprising, as the assumptions that have gone into our derivation are subtly different than the standard assumptions about metallic transport.   In particular, the ordinary derivation of the Wiedemann-Franz law assumes that elastic scattering off of all disorder is much faster than any thermalization time scale (hence, the conductivity can be written as the sum of conductivities of quasiparticles at each energy scale \cite{ashcroft}).   In our derivation, we have assumed that intravalley thermalization is much faster than intervalley scattering,  which may be the case when the dominant source of disorder is long wavelength \cite{lucas, lucas3}.   At a technical level, the integral in the numerator of $\sigma^{\mathrm{anom}}_{zz}$ looks much like the standard integral for $\bar\kappa_{zz}$ in a metal,   and vice versa.    The Wiedemann-Franz law is restored by a factor of $(\mpi^2/3)^2$ coming from the ratio $(2G/C)^2$.    That the Wiedemann-Franz law can arise in a subtle way is emphasized by our interesting violation of the standard Mott relation, which states that $\alpha_{zz} = -(\mpi^2 T/3) (\partial \sigma_{zz}(T=0)/\partial\mu)$.   This Mott relation differs by a minus sign from the result derived above.   The origin of this minus sign is that in our theory, the rate of intervalley scattering is the sum of rates at each quasiparticle energy, as opposed to the net conductivity.

\begin{figure}[t]
\centering
\includegraphics[width=3.5in]{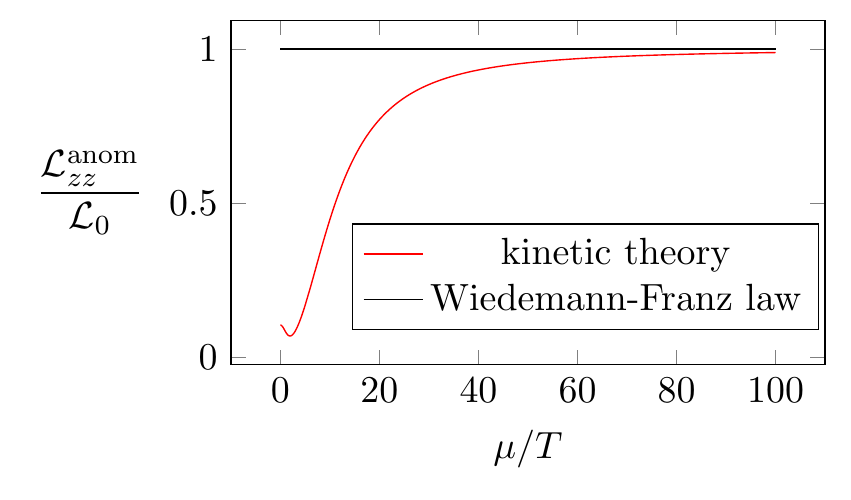}
\caption{Breakdown of the anomalous Wiedemann-Franz law in the regime where intervalley exchange of charge and energy occurs via quasiparticle scattering and may be treated with kinetic theory.  The violation of the Wiedemann-Franz law is opposite to what would be expected in semiconductors or Dirac semimetals such as graphene.   We have assumed that $W_0$ is a constant, though this plot looks qualitatively similar for other choices.}
\label{fig:weylBD}
\end{figure}

Let us also mention what happens in the regime $\mu \sim T$.    In an ordinary semiconductor \cite{goldsmid}, or a Dirac semimetal such as graphene \cite{yoshino}, kinetic theory predicts an $\mathcal{O}(1)$ violation of the Wiedemann-Franz law whereby $\mathcal{L}_{zz} > \mathcal{L}_0$.  This is called bipolar diffusion, and is due to the fact that multiple bands with opposite charge carriers are thermally populated, and the thermal conductivity is enhanced by the combined flow of these carriers.    Figure \ref{fig:weylBD} shows the fate of the anomalous Wiedemann-Franz law in a Weyl semimetal where intervalley scattering is the slowest timescale in the problem.    Here we see the opposite effect -- the Wiedemann-Franz law is reduced.   The physical explanation of this effect immediately follows from the previous paragraph -- bipolar diffusion applies to the scattering rates and not to the conductivities, and hence $\kappa_{zz}$ is reduced below $\sigma_{zz}$ as $\mu/T \rightarrow 0$.   This discussion should be taken with a grain of salt -- it is worth keeping in mind that the regime $\mu/T \rightarrow 0$ is associated with stronger interactions, and so (as in graphene) the quasiparticle description of transport may completely breakdown \cite{crossno, lucas3}.

\section{Memory Matrix Formalism}\label{sec4}
So far, our theory of transport has relied entirely on a classical theory of anomalous hydrodynamics.   Nonetheless, we expect that our results can be computed perturbatively using a more general, inherently quantum mechanical formalism called the memory matrix formalism \cite{forster, lucasMM}.   The memory matrix formalism is an old many-body approach to transport which does not rely on the existence of long-lived quasiparticles.   It is particularly useful in a ``hydrodynamic" regime in which only a small number of quantities are long-lived.   In such a regime, memory matrix results can be understood for many purposes entirely from classical hydrodynamics \cite{lucas}. Nevertheless, the memory matrix formalism has some distinct advantages. In particular, it gives microscopic relations for the unknown parameters of the hydrodynamic theory.

Let us give a simple example of how the memory matrix formalism works, leaving technical details to \cite{forster, lucasMM}.  Suppose we have a system in which the momentum operator $P_i$ is almost exactly conserved. Assuming isotropy, and that there are no other long-lived vector operators, it can be formally shown that the expectation value of $P_i$ will evolve according according to
\begin{equation}
\frac{\mathrm{d} \langle P_i\rangle}{\mathrm{d}t} = - \frac{M_{PP}}{\chi_{PP}} \langle P_i \rangle,  \label{eq:mem1} 
\end{equation}
where $\chi_{PP} = \mathrm{Re}(G^{\mathrm{R}}_{P_x P_x}(\mathbf{k}=\mathbf{0},\omega=0))$ is the momentum-momentum susceptibility, and $M_{PP}$ is a component of the memory matrix, which is schematically given by
\begin{equation}
M_{PP} \approx \lim_{\omega\rightarrow 0} \frac{\mathrm{Im}\left(G^{\mathrm{R}}_{\dot{P}_x \dot{P}_x}(\mathbf{k}=\mathbf{0},\omega)\right)}{\omega}.  \label{eq:mem2}
\end{equation}
More formal expressions may be found in \cite{lucasMM, davison1602}.   Note the presence of operator time derivatives (i.e. $\dot{P} = \mathrm{i}[H,P]$, with $H$ the global Hamiltonian) in the expression for $M_{PP}$.  From the hydrodynamic equation (\ref{eq:mem1}), it is clear that the momentum relaxation rate is determined by $M_{PP}$. For a given microscopic Hamiltonian $H$, we can therefore simply evaluate this element of the memory matrix to obtain the value of the momentum relaxation rate in the hydrodynamic theory.

The memory matrix formalism is very naturally suited to the computation of our hydrodynamic parameters $\mathcal{R}_{ab}$, $\mathcal{S}_{ab}$, $\mathcal{U}_{ab}$ and $\mathcal{V}_{ab}$. As these only affect the conductivities at $\mathcal{O}(B^2)$, it is sufficient to evaluate these in the $B=0$ state. We assume that we may cleanly divide up the low energy effective theory for our Weyl semimetal into ``node fluids" labeled by indices $a$, just as in the main text.   To each node fluid, we assign a charge current operator $J^\mu_a$ and a stress tensor $T^{\mu\nu}_a$, which need not be exactly conserved in the presence of intervalley scattering and anomalies.   We then define the valley charge and energy \emph{operators} as \begin{subequations}\label{eq:opeq}\begin{align}
n_a &\equiv \frac{1}{V_3}\int \mathrm{d}^3\mathbf{x}\; J^t_a, \\
\epsilon_a &\equiv \frac{1}{V_3} \int \mathrm{d}^3\mathbf{x}\; T^{tt}_a,
\end{align}\end{subequations}
respectively.   We have assumed that the fluid is at rest when deriving the above.   For later reference, we also define $\mathcal{J}^i_{a}$ as the zero mode of the operator $J^i_a$, and $P^i_a$ as the zero mode of the operator $T^{ti}_a$, analogous to (\ref{eq:opeq}).

Now suppose that we take our Weyl semimetal, and ``populate" valley fluids at various chemical potentials and temperatures.   Let us define the vector of operators \begin{equation}
x_I = \left(\begin{array}{c} n_a  - n_a^0\\ \epsilon_a - \epsilon_a^0\end{array}\right),
\end{equation} where $n_a^0 = \langle n_a\rangle $ and $\epsilon_a^0 = \langle \epsilon_a\rangle$, with averages over quantum and thermal fluctuations taken in equilibrium.    $I$ indices run over the operators $n_a$ and $\epsilon_a$. Assuming that there are no other long-lived modes operators in the system which overlap with the charge and energy of each valley fluid, we can use memory matrix techniques to show that the expectations values of these objects will evolve according to the hydrodynamic equations  
\begin{equation}
\label{eq:memmatrixhydro}
\frac{\mathrm{d} \langle x_I \rangle }{\mathrm{d}t}   = -M_{IJ}\chi^{-1}_{JK} \langle x_K \rangle,  
\end{equation}
where the matrices $M$ and $\chi$ have entries
\begin{subequations}\label{eq:mem3}\begin{align}
M_{IJ} &\approx \lim_{\omega\rightarrow 0} \frac{\mathrm{Im}\left(G^{\mathrm{R}}_{\dot{x}_I \dot{x}_J}(\mathbf{k}=\mathbf{0},\omega)\right)}{\omega},  \label{eq:memdef}\\
\chi_{IJ} &=  \mathrm{Re}\left(G^{\mathrm{R}}_{x_Ix_J}(\mathbf{k}=\mathbf{0}, \omega=0)\right).
\end{align}\end{subequations}
These formulae should be valid to leading order in a perturbative expansion in the small intervalley coupling strength.
 
The easiest way to compute $\chi_{IJ}$ is to identify the thermodynamic conjugate variable to $x_I$ (let us call it $y_I$), and then employ the linear response formula \begin{equation}
\frac{\partial \langle x_I\rangle}{\partial y_J} = \chi_{IJ}.  \label{eq:LR}
\end{equation}
If the valley fluids interact weakly then we may approximate $\chi_{JK}$ as a block diagonal matrix to leading order, with $T\nu_a$ the canonically conjugate variable to $\mu_a$, and $-T\beta_a$ the canonically conjugate variable to $\epsilon_a$. Thus
 \begin{equation}
\left(\begin{array}{c} \langle n_a\rangle - n_a^0   \\ \langle \epsilon_a \rangle - \epsilon^0_a  \end{array}\right)  = \chi_{IJ} \left(\begin{array}{c} T(\nu_a - \nu_a^0)   \\ -T(\beta_a - \beta_a^0) \end{array}\right).  \label{eq:chi}
\end{equation}
Comparing (\ref{eq:memmatrixhydro}), (\ref{eq:chi}) and our hydrodynamic definition of $\mathsf{A}$, we conclude that $A_{IJ}$, the elements of the intervalley scattering matrix, are related to microscopic Green's functions by\begin{equation}
A_{IJ}  = T M_{IJ}.
\end{equation}
Using the symmetry properties of Green's functions, we see that $M_{IJ}=M_{JI}$, thus proving that $\mathsf{A}$ is a symmetric matrix, as we claimed previously.   From (\ref{eq:mem3}), it is clear that global charge and energy conservation among all valleys enforces $\sum_b \mathcal{R}_{ab} = \sum_b \mathcal{S}_{ab} = \sum_b \mathcal{V}_{ab} = 0$  in the memory matrix formalism. For completeness, we note that the susceptibility matrix is given by
\begin{align}
(\chi_{IJ})_{a\text{ indices}} &= \left(\begin{array}{cc}  (\partial_\mu n)_a &\ 3n_a \\ 3n_a &\ 12P_a \end{array}\right),
\end{align}
assuming that the free energy of each fluid depends only on $\mu_a$ and $T_a$.

We finish by reviewing the well-known microscopic expressions for the other parameters in our hydrodynamic theory (see \cite{lucasMM} for more details). Using the fact that velocity is conjugate to momentum, and combining (\ref{eq:5b}) and (\ref{eq:LR}), we obtain \begin{subequations}\begin{align}
n_a \mdelta^{ij} &\equiv  \chi_{\mathcal{J}^i_{a}P^j_{a}}, \\
(\epsilon_a+P_a)\mdelta^{ij} &\equiv \chi_{P^i_{a}P^j_{a}}.
\end{align}\end{subequations}
The Gibbs-Duhem relation implies that (to good approximation if valley fluids nearly decouple) \begin{equation}
Ts_a = \epsilon_a+P_a - \mu n_a.
\end{equation}
Together with the memory matrix result for the momentum relaxation time \begin{equation}
M_{P^i_{a}P^j_{a}} = \mdelta_{ij} \Gamma_a,
\end{equation}
we have a microscopic expression for all of the hydrodynamic parameters in our formulas for the conductivities, written in the main text, via the memory matrix formalism. The expression (\ref{eq:gammahydro}) for $\Gamma_a$ that we derived from hydrodynamics agrees with that obtained by explicitly evaluating $M_{PP}$ \cite{dsz}.

It is possible that the presence of anomalies complicates the memory matrix formalism beyond what is anticipated above.   However, as the anomalous contributions to the hydrodynamic equations vanish in the absence of external electromagnetic fields,  we do not expect any difficulties when the memory matrices are computed in the absence of background magnetic fields. 

  \end{appendix}

\addcontentsline{toc}{section}{References}
\bibliographystyle{unsrt}
\bibliography{theorypaperbib}

\begin{thebibliography}{10}

\bibitem{vishwanath}
X.~Wan, A.~M. Turner, A.~Vishwanath, and S.~Y. Savrasov.
\newblock ``Topological semimetal and Fermi-arc surface states in the
  electronic structure of pyrochlore iridates",
  \href{http://journals.aps.org/prb/abstract/10.1103/PhysRevB.83.205101}{\textsl{Physical
  Review} \textbf{B83} \texttt{205101} (2011)}.

\bibitem{bernevig1}
C.~Fang, M.~J. Gilbert, X.~Dai, and A.~Bernevig.
\newblock ``Multi-Weyl topological semimetals stabilized by point group
  symmetry",
  \href{http://journals.aps.org/prl/abstract/10.1103/PhysRevLett.108.266802}{\textsl{Physical
  Review Letters} \textbf{108} \texttt{266802} (2012)},
  \href{http://arxiv.org/abs/1111.7309}{\texttt{arXiv:1111.7309}}.

\bibitem{bernevig2}
H.~Weng, C.~Fang, Z.~Fang, A.~Bernevig, and X.~Dai.
\newblock ``Weyl semimetal phase in non-centrosymmetric transition metal
  monophosphides",
  \href{http://journals.aps.org/prx/abstract/10.1103/PhysRevX.5.011029}{\textsl{Physical
  Review} \textbf{X5} \texttt{011029} (2015)},
  \href{http://arxiv.org/abs/1501.00060}{\texttt{arXiv:1501.00060}}.

\bibitem{soljacic}
L.~Lu, Z.~Wang, D.~Ye, L.~Ran, L.~Fu, J.~D. Joannoupoulos, and
  M.~Solja\v{c}i\'c.
\newblock ``Experimental observation of Weyl points",
  \href{http://www.sciencemag.org/content/349/6248/622}{\textsl{Science}
  \textbf{349} 622 (2015)},
  \href{http://arxiv.org/abs/1502.03438}{\texttt{arXiv:1502.03438}}.

\bibitem{syxu}
S.-Y.~Xu \emph{et al.}
\newblock ``Discovery of a Weyl fermion semimetal and topological Fermi arcs",
  \href{http://www.sciencemag.org/content/349/6248/613}{\textsl{Science}
  \textbf{349} 613 (2015)},
  \href{http://arxiv.org/abs/1502.03807}{\texttt{arXiv:1502.03807}}.

\bibitem{bqlv}
B.~Q.~Lv \emph{et al.}
\newblock ``Experimental discovery of Weyl semimetal TaAs",
  \href{http://journals.aps.org/prx/abstract/10.1103/PhysRevX.5.031013}{\textsl{Physical
  Review} \textbf{X5} \texttt{031013} (2015)},
  \href{http://arxiv.org/abs/1502.04684}{\texttt{arXiv:1502.04684}}.

\bibitem{nielsen}
H.~N. Nielsen and M.~Ninomiya.
\newblock ``The Adler-Bell-Jackiw anomaly and Weyl fermions in a crystal",
  \href{http://www.sciencedirect.com/science/article/pii/0370269383915290}{\textsl{Physics
  Letters} \textbf{B130} 389 (1983)}.

\bibitem{spivakson}
D.~T. Son and B.~Z. Spivak.
\newblock ``Chiral anomaly and classical negative magnetoresistance of Weyl
  metals",
  \href{http://journals.aps.org/prb/abstract/10.1103/PhysRevB.88.104412}{\textsl{Physical
  Review} \textbf{B88} \texttt{104412} (2013)},
  \href{http://arxiv.org/abs/1206.1627}{\texttt{arXiv:1206.1627}}.

\bibitem{burkov}
A.~A. Burkov.
\newblock ``Negative longitudinal magnetoresistance in Dirac and Weyl metals",
  \href{http://journals.aps.org/prb/abstract/10.1103/PhysRevB.91.245157}{\textsl{Physical
  Review} \textbf{B91} \texttt{245157} (2015)},
  \href{http://arxiv.org/abs/1505.01849}{\texttt{arXiv:1505.01849}}.

\bibitem{fiete}
R.~Lundgren, P.~Laurell, and G.~A. Fiete.
\newblock ``Thermoelectric properties of Weyl and Dirac semimetals",
  \href{http://journals.aps.org/prb/abstract/10.1103/PhysRevB.90.165115}{\textsl{Physical
  Review} \textbf{B90} \texttt{165115} (2014)},
  \href{http://arxiv.org/abs/1407.1435}{\texttt{arXiv:1407.1435}}.

\bibitem{spivak1510}
B.~Z. Spivak and A.~V. Andreev.
\newblock ``Magneto-transport phenomena related to the chiral anomaly in Weyl
  semimetals",
  \href{http://journals.aps.org/prb/abstract/10.1103/PhysRevB.93.085107}{\textsl{Physical
  Review} \textbf{B93} \texttt{085107} (2016)},
  \href{http://arxiv.org/abs/1510.01817}{\texttt{arXiv:1510.01817}}.

\bibitem{hjkim}
H-J. Kim, K-S. Kim, J-F. Wang, M.~Sasaki, N.~Satoh, A.~Ohnishi, M.~Kitaura,
  M.~Yang, and L.~Li.
\newblock ``Dirac vs. Weyl in topological insulators: Adler-Bell-Jackiw anomaly
  in transport phenomena",
  \href{http://journals.aps.org/prl/abstract/10.1103/PhysRevLett.111.246603}{\textsl{Physical
  Review Letters} \textbf{111} \texttt{246603} (2013)},
  \href{http://arxiv.org/abs/1307.6990}{\texttt{arXiv:1307.6990}}.

\bibitem{xiong}
J.~Xiong, S.~K. Kushwaha, T.~Liang, J.~W. Krizan, M.~Hirschberger, W.~Wang,
  R.~J. Cava, and N.~P. Ong.
\newblock ``Evidence for the chiral anomaly in the Dirac semimetal
  $\mathrm{Na}_3\mathrm{Bi}$",
  \href{http://science.sciencemag.org/content/350/6259/413}{\textsl{Science}
  \textbf{350} 413 (2015)}.

\bibitem{huangprx}
X.~Huang \emph{et al.}
\newblock ``Observation of the chiral anomaly induced negative
  magnetoresistance in 3D Weyl semimetal TaAs",
  \href{http://journals.aps.org/prx/abstract/10.1103/PhysRevX.5.031023}{\textsl{Physical
  Review} \textbf{X5} \texttt{031023} (2015)},
  \href{http://arxiv.org/abs/1503.01304}{\texttt{arXiv:1503.01304}}.

\bibitem{czli}
C-Z. Li, L-X. Wang, H.~Liu, J.~Wang, Z-M. Lin, and D-P. Yu.
\newblock ``Giant negative magnetoresistance induced by the chiral anomaly in
  individual $\mathrm{Cd}_3\mathrm{As}_2$ nanowires",
  \href{http://www.nature.com/ncomms/2015/151217/ncomms10137/full/ncomms10137.html}{\textsl{Nature
  Communications} \textbf{6} \texttt{10137} (2015)},
  \href{http://arxiv.org/abs/1504.07398}{\texttt{arXiv:1504.07398}}.

\bibitem{hli}
H.~Li, H.~He, H-Z. Lu, H.~Zhang, H.~Liu, R.~Ma, Z.~Fan, S-Q. Shen, and J.~Wang.
\newblock ``Negative magnetoresistance in Dirac semimetal
  $\mathrm{Cd}_3\mathrm{As}_2$",
  \href{http://www.nature.com/ncomms/2016/160108/ncomms10301/full/ncomms10301.html}{\textsl{Nature
  Communications} \textbf{7} \texttt{10301} (2016)},
  \href{http://arxiv.org/abs/1507.06470}{\texttt{arXiv:1507.06470}}.

\bibitem{czhang}
C.~Zhang \emph{et al.}
\newblock ``Signatures of the Adler-Bell-Jackiw chiral anomaly in a Weyl
  fermion semimetal",
  \href{http://www.nature.com/ncomms/2016/160225/ncomms10735/full/ncomms10735.html}{\textsl{Nature
  Communications} \textbf{7} \texttt{10735} (2016)},
  \href{http://arxiv.org/abs/1601.04208}{\texttt{arXiv:1601.04208}}.

\bibitem{hirschberger}
M.~Hirschberger, S.~Kushwaha, Z.~Wang, Q.~Gibson, C.~A. Belvin, B.~A. Bernevig,
  R.~J. Cava, and N.~P. Ong.
\newblock ``The chiral anomaly and thermopower of Weyl fermions in the
  half-Heusler GdPtBi",
  \href{http://arxiv.org/abs/1602.07219}{\texttt{arXiv:1602.07219}}.

\bibitem{gurzhi1}
R.~N. Gurzhi.
\newblock ``Minimum of resistance in impurity-free conductors",
  \href{http://www.jetp.ac.ru/cgi-bin/e/index/e/17/2/p521?a=list}{\textsl{Journal
  of Experimental and Theoretical Physics} \textbf{17} 521 (1963)}.

\bibitem{spivak2006}
B.~Spivak and S.~A. Kivelson.
\newblock ``Transport in two-dimensional electronic micro-emulsions",
  \href{http://www.sciencedirect.com/science/article/pii/S0003491605002654}{\textsl{Annals
  of Physics} \textbf{321} 2079 (2006)}.

\bibitem{andreev}
A.~V. Andreev, S.~A. Kivelson, and B.~Spivak.
\newblock ``Hydrodynamic description of transport in strongly correlated
  electron systems",
  \href{http://journals.aps.org/prl/abstract/10.1103/PhysRevLett.106.256804}{\textsl{Physical
  Review Letters} \textbf{106} \texttt{256804} (2011)},
  \href{http://arxiv.org/abs/1011.3068}{\texttt{arXiv:1011.3068}}.

\bibitem{tomadin}
A.~Tomadin, G.~Vignale, and M.~Polini.
\newblock ``A Corbino disk viscometer for 2d quantum electron liquids",
  \href{http://journals.aps.org/prl/abstract/10.1103/PhysRevLett.113.235901}{\textsl{Physical
  Review Letters} \textbf{113} \texttt{235901} (2014)},
  \href{http://arxiv.org/abs/1401.0938}{\texttt{arXiv:1401.0938}}.

\bibitem{vignale}
A.~Principi and G.~Vignale.
\newblock ``Violation of the Wiedemann-Franz law in hydrodynamic electron
  liquids",
  \href{http://journals.aps.org/prl/abstract/10.1103/PhysRevLett.115.056603}{\textsl{Physical
  Review Letters} \textbf{115} \texttt{056603} (2015)}.

\bibitem{polini}
I.~Torre, A.~Tomadin, A.~K. Geim, and M.~Polini.
\newblock ``Nonlocal transport and the hydrodynamic shear viscosity in
  graphene",
  \href{http://journals.aps.org/prb/abstract/10.1103/PhysRevB.92.165433}{\textsl{Physical
  Review} \textbf{B92} \texttt{165433} (2015)},
  \href{http://arxiv.org/abs/1508.00363}{\texttt{arXiv:1508.00363}}.

\bibitem{levitovhydro}
L.~Levitov and G.~Falkovich.
\newblock ``Electron viscosity, current vortices and negative nonlocal
  resistance in graphene",
  \href{http://arxiv.org/abs/1508.00836}{\texttt{arXiv:1508.00836}}.

\bibitem{molenkamp}
M.~J.~M. de~Jong and L.~W. Molenkamp.
\newblock ``Hydrodynamic electron flow in high-mobility wires",
  \href{http://journals.aps.org/prb/abstract/10.1103/PhysRevB.51.13389}{\textsl{Physical
  Review} \textbf{B51} 11389 (1995)},
  \href{http://arxiv.org/abs/cond-mat/9411067}{\texttt{arXiv:cond-mat/9411067}}.

\bibitem{bandurin}
D.~A.~Bandurin \emph{et al.}
\newblock ``Negative local resistance due to viscous electron backflow in
  graphene",
  \href{http://science.sciencemag.org/content/351/6277/1055.long}{\textsl{Science}
  \textbf{351} 1055 (2016)},
  \href{http://arxiv.org/abs/1509.04165}{\texttt{arXiv:1509.04165}}.

\bibitem{mackenzie}
P.~J.~W. Moll, P.~Kushwaha, N.~Nandi, B.~Schmidt, and A.~P. Mackenzie.
\newblock ``Evidence for hydrodynamic electron flow in $\mathrm{PdCoO}_2$",
  \href{http://science.sciencemag.org/content/351/6277/1061}{\textsl{Science}
  \textbf{351} 1061 (2016)},
  \href{http://arxiv.org/abs/1509.05691}{\texttt{arXiv:1509.05691}}.

\bibitem{crossno}
J.~Crossno \emph{et al.}
\newblock ``Observation of the Dirac fluid and the breakdown of the
  Wiedemann-Franz law in graphene",
  \href{http://science.sciencemag.org/content/351/6277/1058}{\textsl{Science}
  \textbf{351} 1058 (2016)},
  \href{http://arxiv.org/abs/1509.04713}{\texttt{arXiv:1509.04713}}.

\bibitem{ghaharikim}
F.~Ghahari, H-Y. Xie, T.~Taniguchi, K.~Watanabe, M.~S. Foster, and P.~Kim.
\newblock ``Enhanced Thermoelectric Power in Graphene: Violation of the Mott
  Relation By Inelastic Scattering",
  \href{http://journals.aps.org/prl/abstract/10.1103/PhysRevLett.116.136802}{\textsl{Physical
  Review Letters} \textbf{116} \texttt{136802} (2016)},
  \href{http://arxiv.org/abs/1601.05859}{\texttt{arXiv:1601.05859}}.

\bibitem{hkms}
S.~A. Hartnoll, P.~K. Kovtun, M.~M\"uller, and S.~Sachdev.
\newblock ``Theory of the Nernst effect near quantum phase transitions in
  condensed matter, and in dyonic black holes",
  \href{http://journals.aps.org/prb/abstract/10.1103/PhysRevB.76.144502}{\textsl{Physical
  Review} \textbf{B76} \texttt{144502} (2007)},
  \href{http://arxiv.org/abs/0706.3215}{\texttt{arXiv:0706.3215}}.

\bibitem{lucas3}
A.~Lucas, J.~Crossno, K.~C. Fong, P.~Kim, and S.~Sachdev.
\newblock ``Transport in inhomogeneous quantum critical fluids and in the Dirac
  fluid in graphene",
  \href{http://journals.aps.org/prb/abstract/10.1103/PhysRevB.93.075426}{\textsl{Physical
  Review} \textbf{B93} \texttt{075426} (2016)},
  \href{http://arxiv.org/abs/1510.01738}{\texttt{arXiv:1510.01738}}.

\bibitem{lucasplasma}
A.~Lucas.
\newblock ``Sound waves and resonances in electron-hole plasma",
  \href{http://arxiv.org/abs/1604.03955}{\texttt{arXiv:1604.03955}}.

\bibitem{surowka}
D.~T. Son and P.~Sur\'owka.
\newblock ``Hydrodynamics with triangle anomalies",
  \href{http://journals.aps.org/prl/abstract/10.1103/PhysRevLett.103.191601}{\textsl{Physical
  Review Letters} \textbf{103} \texttt{191601} (2009)},
  \href{http://arxiv.org/abs/0906.5044}{\texttt{arXiv:0906.5044}}.

\bibitem{oz}
Y.~Neiman and Y.~Oz.
\newblock ``Relativistic hydrodynamics with general anomalous charges",
  \href{http://link.springer.com/article/10.1007%2FJHEP03%282011%29023}{\textsl{Journal
  of High Energy Physics} \textbf{03} \texttt{023} (\textbf{2011})},
  \href{http://arxiv.org/abs/1011.5107}{\texttt{arXiv:1011.5107}}.

\bibitem{landsteinerhydro}
K.~Landsteiner, Y.~Liu, and Y-W. Sun.
\newblock ``Negative magnetoresistivity in chiral fluids and holography",
  \href{http://link.springer.com/article/10.1007%2FJHEP09%282015%29090}{\textsl{Journal
  of High Energy Physics} \textbf{03} \texttt{127} (\textbf{2015})},
  \href{http://arxiv.org/abs/1410.6399}{\texttt{arXiv:1410.6399}}.

\bibitem{Sadofyev:2010pr}
A.~V. Sadofyev and M.~V. Isachenkov.
\newblock ``The chiral magnetic effect in hydrodynamical approach",
  \href{http://www.sciencedirect.com/science/article/pii/S0370269311001869}{\textsl{Physics
  Letters} \textbf{B697} 404 (2011)},
  \href{http://arxiv.org/abs/1010.1550}{\texttt{arXiv:1010.1550}}.

\bibitem{Roychowdhury:2015jha}
D.~Roychowdhury.
\newblock ``Magnetoconductivity in chiral Lifshitz hydrodynamics",
  \href{http://link.springer.com/article/10.1007%2FJHEP09%282015%29145}{\textsl{Journal
  of High Energy Physics} \textbf{09} \texttt{145} (\textbf{2015})},
  \href{http://arxiv.org/abs/1508.02002}{\texttt{arXiv:1508.02002}}.

\bibitem{lucasMM}
A.~Lucas and S.~Sachdev.
\newblock ``Memory matrix theory of magnetotransport in strange metals",
  \href{http://journals.aps.org/prb/abstract/10.1103/PhysRevB.91.195122}{\textsl{Physical
  Review} \textbf{B91} \texttt{195122} (2015)},
  \href{http://arxiv.org/abs/1502.04704}{\texttt{arXiv:1502.04704}}.

\bibitem{jensen}
K.~Jensen, R.~Loganayagam, and A.~Yarom.
\newblock ``Thermodynamics, gravitational anomalies and cones",
  \href{http://link.springer.com/article/10.1007%2FJHEP03%282011%29023}{\textsl{Journal
  of High Energy Physics} \textbf{02} \texttt{088} (\textbf{2013})},
  \href{http://arxiv.org/abs/1207.5824}{\texttt{arXiv:1207.5824}}.

\bibitem{landsteiner}
K.~Landsteiner.
\newblock ``Anomaly related transport of Weyl fermions for Weyl semi-metals",
  \href{http://journals.aps.org/prb/abstract/10.1103/PhysRevB.89.075124}{\textsl{Physical
  Review} \textbf{B89} \texttt{075124} (2014)},
  \href{http://arxiv.org/abs/1306.4932}{\texttt{arXiv:1306.4932}}.

\bibitem{Jensen:2012jh}
K.~Jensen, M.~Kaminski, P.~Kovtun, R.~Meyer, A.~Ritz, and A.~Yarom.
\newblock ``Towards hydrodynamics without an entropy current",
  \href{http://journals.aps.org/prl/abstract/10.1103/PhysRevLett.109.101601}{\textsl{Physical
  Review Letters} \textbf{109} \texttt{101601} (2012)},
  \href{http://arxiv.org/abs/1203.3556}{\texttt{arXiv:1203.3556}}.

\bibitem{minwalla}
N.~Banerjee, J.~Bhattacharya, S.~Bhattacharyya, S.~Jain, S.~Minwalla, and
  T.~Sharma.
\newblock ``Constraints on fluid dynamics from equilibrium partition
  functions",
  \href{http://link.springer.com/article/10.1007%2FJHEP09%282012%29046}{\textsl{Journal
  of High Energy Physics} \textbf{09} \texttt{046} (\textbf{2012})},
  \href{http://arxiv.org/abs/1203.3544}{\texttt{arXiv:1203.3544}}.

\bibitem{landaustat}
L.~D. Landau and E.~M. Lifshitz.
\newblock \emph{Statistical Physics Part 1}
  \href{http://www.amazon.com/Statistical-Physics-Third-Part-Theoretical/dp/0750633727/ref=sr_1_1?ie=UTF8&qid=1460470916&sr=8-1&keywords=landau+statistical+physics}{(Butterworth
  Heinemann, $3^{\mathrm{rd}}$ ed., 1980)}.

\bibitem{forster}
D.~Forster.
\newblock \emph{Hydrodynamic Fluctuations, Broken Symmetry and Correlation
  Functions}
  \href{http://www.amazon.com/Hydrodynamic-Fluctuations-Symmetry-Correlation-Functions/dp/0201410494/ref=sr_1_1?ie=UTF8&qid=1418231940&sr=8-1&keywords=hydrodynamic+fluctuations+broken+symmetry+and+correlation+functions&pebp=1418231941724}{(Perseus
  Books, 1975)}.

\bibitem{lucas}
A.~Lucas.
\newblock ``Hydrodynamic transport in strongly coupled disordered quantum field
  theories",
  \href{http://iopscience.iop.org/article/10.1088/1367-2630/17/11/113007/meta}{\textsl{New
  Journal of Physics} \textbf{17} \texttt{113007} (2015)},
  \href{http://arxiv.org/abs/1506.02662}{\texttt{arXiv:1506.02662}}.

\bibitem{kharzeevreview}
D.~E. Kharzeev.
\newblock ``The chiral magnetic effect and anomaly-induced transport",
  \href{http://www.sciencedirect.com/science/article/pii/S0146641014000039}{\textsl{Progress
  in Particle and Nuclear Physics} \textbf{75} 133 (2014)},
  \href{http://arxiv.org/abs/1312.3348}{\texttt{arXiv:1312.3348}}.

\bibitem{vazifeh}
M.~M. Vazifeh and M.~Franz.
\newblock ``Electromagnetic response of Weyl semimetals",
  \href{http://journals.aps.org/prl/abstract/10.1103/PhysRevLett.111.027201}{\textsl{Physical
  Review Letters} \textbf{111} \texttt{027201} (2013)},
  \href{http://arxiv.org/abs/1303.5784}{\texttt{arXiv:1303.5784}}.

\bibitem{hartnollads}
S.~A. Hartnoll.
\newblock ``Lectures on holographic methods for condensed matter physics",
  \href{http://iopscience.iop.org/0264-9381/26/22/224002/}{\textsl{Classical
  and Quantum Gravity} \textbf{26} \texttt{224002} (2009)},
  \href{http://arxiv.org/abs/0903.3246}{\texttt{arXiv:0903.3246}}.

\bibitem{ashcroft}
N.~W. Ashcroft and N.~D. Mermin.
\newblock \emph{Solid-State Physics}
  \href{http://www.amazon.com/Solid-State-Physics-Neil-Ashcroft/dp/0030839939/ref=sr_1_1?ie=UTF8&qid=1439625478&sr=8-1&keywords=ashcroft+and+mermin}{(Brooks
  Cole, 1976)}.

\bibitem{goldsmid}
G.~S. Nolas and H.~J. Goldsmid.
\newblock ``Thermal conductivity of semiconductors", in \emph{Thermal
  Conductivity: Theory, Properties and Applications} (ed. T. M. Tritt), 105,
  \href{http://www.amazon.com/Quantum-Phase-Transitions-Subir-Sachdev/dp/0521514681/ref=sr_1_1?ie=UTF8&qid=1433031213&sr=8-1&keywords=quantum+phase+transitions}{(Kluwer
  Academic, 2004)}.

\bibitem{landsteiner2}
M.~N. Chernodub, A.~Cortijo, A.~G. Grushin, K.~Landsteiner, and M.~A.~H.
  Vozmediano.
\newblock ``Condensed matter realization of the axial magnetic effect",
  \href{http://journals.aps.org/prb/abstract/10.1103/PhysRevB.89.081407}{\textsl{Physical
  Review} \textbf{B89} \texttt{081407} (2014)},
  \href{http://arxiv.org/abs/1311.0878}{\texttt{arXiv:1311.0878}}.

\bibitem{kaminski}
M.~Kaminski, C.~F. Uhlemann, M.~Bleicher, and J.~Schaffner-Bielich.
\newblock ``Anomalous hydrodynamics kicks neutron stars",
  \href{http://arxiv.org/abs/1410.3833}{\texttt{arXiv:1410.3833}}.

\bibitem{maissam}
R.~Mahajan, M.~Barkeshli, and S.~A. Hartnoll.
\newblock ``Non-Fermi liquids and the Wiedemann-Franz law",
  \href{http://journals.aps.org/prb/abstract/10.1103/PhysRevB.89.075124}{\textsl{Physical
  Review} \textbf{B88} \texttt{125107} (2013)},
  \href{http://arxiv.org/abs/1304.4249}{\texttt{arXiv:1304.4249}}.

\bibitem{yoshino}
H.~Yoshino and K.~Murata.
\newblock ``Significant enhancement of electronic thermal conductivity of
  two-dimensional zero-gap systems by bipolar-diffusion effect",
  \href{http://journals.jps.jp/doi/abs/10.7566/JPSJ.84.024601}{\textsl{Journal
  of the Physical Society of Japan} \textbf{84} \texttt{024601} (2015)}.

\bibitem{davison1602}
R.~A. Davison, L.~V. Delacr\'etaz, B.~Gout\'eraux, and S.~A. Hartnoll.
\newblock ``Hydrodynamic theory of quantum fluctuating superconductivity",
  \href{http://arxiv.org/abs/1602.08171}{\texttt{arXiv:1602.08171}}.

\bibitem{dsz}
R.~A. Davison, K.~Schalm, and J.~Zaanen.
\newblock ``Holographic duality and the resistivity of strange metals",
  \href{http://journals.aps.org/prb/abstract/10.1103/PhysRevB.89.245116}{\textsl{Physical
  Review} \textbf{B89} \texttt{245116} (2014)},
  \href{http://arxiv.org/abs/1311.2451}{\texttt{arXiv:1311.2451}}.

\end{thebibliography}
\end{document}